\documentclass[useAMS,usenatbib]{mn2e}

\usepackage{graphicx}
\usepackage{lscape}

\title[Fifteen new T dwarfs in the LAS]{Fifteen new T dwarfs discovered in the UKIDSS Large Area Survey}
\author[D. J. Pinfield et al.]
{D. J. Pinfield$^{1}$\thanks{E-mail: D.J.Pinfield@herts.ac.uk},
B. Burningham$^{1}$, M. Tamura$^{2}$, S. K. Leggett$^{3}$, N. Lodieu$^{4}$, 
\newauthor
P. W. Lucas$^{1}$, D.J. Mortlock$^{5}$, S. J. Warren$^{5}$, D. Homeier$^{6}$, M. Ishi$^{7}$, N. R. Deacon$^{8}$, 
\newauthor
R. G. McMahon$^{9}$, P. C. Hewett$^{9}$, M. R. Zapatero Osorio$^{4}$, E. L. Martin$^{4}$, 
\newauthor
H. R. A. Jones$^{1}$, B.P. Venemans$^{9}$, A. Day-Jones$^{1}$, P. D. Dobbie$^{10}$, S. L. Folkes$^{1}$, 
\newauthor
S. Dye$^{11}$, F. Allard$^{12}$, I. Baraffe$^{13}$, D. Barrado y Navascu\'es$^{14}$, S. L. Casewell$^{15}$, 
\newauthor
K. Chiu$^{16}$, G. Chabrier$^{13}$, F. Clarke$^{17}$, S. T. Hodgkin$^{9}$, A. Magazz\`u$^{18}$, 
\newauthor
M. J. McCaughrean$^{16}$, E. Moraux$^{19}$, T. Nakajima$^{2}$, Y. Pavlenko$^{20}$, C. G. Tinney$^{21}$\\
$^{1}$Centre for Astrophysics Research, Science and Technology Research 
Institute, University of Hertfordshire, Hatfield AL10 9AB \\
$^{2}$National Astronomical Observatory, Mitaka, Tokyo 181-8588 \\
$^{3}$Gemini Observatory, 670 N. A'ohoku Place, Hilo, HI 96720, USA \\
$^{4}$Instituto de Astrof\'isica de Canarias, 38200 La Laguna, Spain \\
$^{5}$Astrophysics Group, Imperial College London, Blackett Laboratory, 
      Prince Consort Road, London SW7 2AZ \\
$^{6}$Institut fur Astrophysik, Georg-August-Universitat, Friedrich-Hund-Platz 1, 
      37077 Gottingen, Germany \\
$^{7}$Subaru Telescope, 650 North A'ohoku Place, Hilo, Hi 96720, USA \\
$^{8}$Department of Astrophysics, Faculty of Science, Radboud University 
       Nijmegen, PO Box 9010, 6500 GL Nijmegen, The Netherlands \\
$^{9}$Institute of Astronomy, Madingley Road, Cambridge CB3 0HA, UK \\
$^{10}$Anglo-Australian Observatory, P.O. Box 296, Epping 1710, Australia \\
$^{11}$Cardif University, School of Physics \& Astronomy, Queens Buildings, The Parade, Cardif, CF24 3AA, U.K. \\
$^{12}$Centre de Recherche Astrophysique de Lyon, UMR5574, CNRS, Universite de Lyon, 
      Ecole Normale Superieure, 46 Allee d'Italie, \\
      F-69364 Lyon Cedex 07, France \\
$^{13}$C.R.A.L. (UMR 5574 CNRS), Ecole Normale Superieure, 69364 Lyon Cedex 07, France \\
$^{14}$Laboratorio de Astrof\'isica Espacial y F\'isica Fundamental,
       INTA, P.O. Box 50727, E--2808 Madrid, Spain \\
$^{15}$Department of Physics and Astronomy, University of Leicester,
       University Road, Leicester LE1 7RH, UK \\
$^{16}$School of Physics, University of Exeter, Stocker Road, Exeter EX4 4QL, Devon, UK \\
$^{17}$European Southern Observatory, Alonso de Cordova 3107, Casilla 19001 
       Santiago 19, Chile
}
\begin{document}
\maketitle
\label{firstpage}

\begin{abstract}
We present the discovery of fifteen new T2.5-T7.5 dwarfs (with estimated distances between 
$\sim$24--93pc), identified in the first three main data releases of the UKIRT Infrared Deep 
Sky Survey. This brings the total number of T dwarfs discovered in the Large Area Survey (to 
date) to 28. These discoveries are confirmed by near infrared spectroscopy, from which we derive 
spectral types on the unified scheme of Burgasser et al. (2006). Seven of the new T dwarfs have 
spectral types of T2.5-T4.5, five have spectral types of T5-T5.5, one is a T6.5p, and two 
are T7-7.5. We assess spectral morphology and colours to identify T dwarfs in our sample 
that may have non-typical physical properties (by comparison to solar neighbourhood populations), 
and find that one of these new T dwarfs may be metal poor, three may have low surface gravity, 
and one may have high surface gravity. The colours of the full sample of LAS T dwarfs show a 
possible trend to bluer $Y-J$ with decreasing effective temperature, and some interesting 
colour changes in $J-H$ and $z-J$ (deserving further investigation) beyond T8. The LAS T dwarf 
sample from the first and second main data releases show good evidence for a consistent level 
of completion to J=19. By accounting for the main sources of incompleteness (selection, follow-up 
and spatial) as well as the effects of unresolved binarity and Malmquist bias, we estimate that 
there are 17$\pm$4 $\ge$T4 dwarfs in the $J\le$19 volume of the LAS second data release. Comparing 
this to theoretical predictions is most consistent with a sub-stellar mass function exponent 
$\alpha$ between -1.0 and 0. This is consistent with the latest 2MASS/SDSS constraint (which 
is based on lower number statistics), and is significantly lower than the $\alpha\sim1.0$ 
suggested by L dwarf field populations, possibly a result of the lower mass range probed by 
the T dwarf class.
\end{abstract}


\section{Introduction}

The advent of the large scale Sloan Digital Sky Survey (SDSS; York et al. 2000), 
the 2-Micron All Sky Survey (2MASS; Skrutskie et al. 2006) and Deep Near Infrared 
Survey of the Southern Sky (DENIS; Epchtein et al. 1997) were the main factors 
that led to the identification and study of the two spectral classes beyond M; 
the L dwarfs (with effective temperature $T_{\rm eff}\sim$2300-1450K; Golimowski 
et al. 2004) have dusty atmospheres and very red near infrared colours, while 
the even cooler T dwarfs ($T_{\rm eff}<$1450K) have clear atmospheres (from 
which the dust has settled), and their blue infrared spectra are dominated 
by strong CH$_4$ and H$_2$O absorption bands. The spectral typing scheme of 
Kirkpatrick et al. (1999) established the first properly classified L dwarf 
population via 2MASS discoveries. The first T dwarf was discovered as a close 
companion to the early M dwarf GL229 (Nakajima et al. 1995), and the first 
L dwarf as a companion to the white dwarf GD 165 (Becklin \& Zuckerman 1988; 
Kirkpatrick et al. 1993). However, the subsequent growth of the known T (and L) 
dwarf populations ($>$500 L dwarfs and $>$100 T dwarfs are known at time of 
writing; Kirkpatrick 2005) has been predominantly achieved via searches of 
the SDSS (e.g. Leggett et al. 2000; Geballe et al. 2002; Knapp et al. 2004; 
Chiu et al. 2006) and 2MASS (e.g. Kirkpatrick et al. 2000; Cruz et al. 2007; 
Burgasser et al. 2002; 2004; Tinney et al. 2005; Looper Kirkpatrick \& Burgasser 
2007) databases.

The coolest $T_{\rm eff}$ probed by SDSS and 2MASS are currently defined 
by the eight T7.5-8 dwarfs (as typed by the Burgasser et al. (2006b; B06) scheme) 
discovered in these surveys. These objects have $T_{\rm eff}$ in the range 725-950K 
(Geballe et al. 2001; Saumon et al. 2006; 2007; Leggett et al. 2007; Burgasser, 
Burrows \& Kirkpatrick 2006). At lower $T_{\rm eff}$ it is possible that a new 
spectral class (pre-emptively called Y, following Kirkpatrick et al. (1999)) 
may be necessary, if for example, spectral changes occur such as the strengthening 
of ammonia absorption, and/or the condensation of atmospheric water clouds 
(Burrows, Sudarsky \& Lunine (2003)). L and T dwarfs already encompass the 
temperature scale of transiting ``hot Jupiters'' (e.g. Knutson et al. 2007), 
and populating the even cooler $T_{\rm eff}$ regime will allow us to study 
and understand atmospheres whose $T_{\rm eff}$ is similar to cooler extra-solar 
giant planets populations.

The current complement of known L and T dwarfs allows some constraints to be 
placed on the substellar mass function (e.g. Chabrier 2003 and references therein; 
Allen et al. 2005; Metchev et al. 2007). However, if the brown dwarf mass function 
is to be accurately constrained, a significantly larger number of late T dwarfs 
will be extremely beneficial, since the $T_{\rm eff}$ distribution of T dwarfs 
in the $<$950K range is particularly sensitive to mass function variations (e.g. 
fig 5 of Burgasser 2004). Also, accurate constraints on the brown dwarf birth 
rate (i.e. their formation history) need the improved statistics that come 
with larger numbers of both L and T dwarfs.

The UKIRT (UK Infrared Telescope) Infrared Deep Sky Survey (UKIDSS; Lawrence 
et al. 2007) is a new infrared survey being conducted with the UKIRT Wide Field 
Camera (WFCAM; Casali et al. 2007). UKIDSS is a set of five sub-surveys, 
with three wide field surveys -- the Large Area Survey (LAS), the Galactic 
Cluster Survey (GCS) and the Galactic Plane Survey (GPS) -- and two 
very deep surveys -- the Deep Extra-galactic Survey (DXS) and the Ultra-Deep 
Survey (UDS). The LAS is the largest of the wide-field surveys, and 
will cover 4000 sq degs of sky in four filter bands, going several magnitudes 
deeper than 2MASS. UKIDSS began in May 2005, and at the time of writing there 
have been four ESO (European Southern Observatory)-wide releases, including 
an Early Data Release (EDR) in February 2006 (Dye et al. 2006), and three 
subsequent main data releases: Data Release 1 (DR1) in 2006 July covering 
190 sq degs (Warren et al. 2007b), Data Release 2 (DR2) in 2007 March covering 
280 sq degs (including DR1; Warren et al. 2007c), and Data Release 3 (DR3) in 
2007 December, covering 900 sq degs (including DR1 and DR2).

These data are providing un-paralleled sensitivity to L and T dwarf 
populations, and our collaboration has begun a variety of UKIDSS-based 
searches for these objects. Here we focus on our search for late T dwarfs 
and potentially new record breaking low-$T_{\rm eff}$ objects. Previous 
LAS T dwarfs have been presented by Kendall et al. (2007), Lodieu et al. 
(2007), Warren et al. (2007a) and Chiu et al. (2008), who have discovered a 
total of 13 spectroscopically confirmed LAS T dwarfs from the EDR and DR1. 
This included the first T8.5 dwarf ULAS J0034-0052 (Warren et al. 2007a).

Here we report the discovery of 15 new LAS T dwarfs discovered in DR1, DR2 
and some of DR3. Section 2 summarises our selection criteria for identifying 
candidate T dwarfs as well as possible 400-700K objects. Section 3 describes 
the follow-up photometry we have obtained to identify spurious objects amongst 
these candidates, and Section 4 presents our spectroscopic follow-up and 
confirmation of the 15 new T dwarfs. Section 5 discusses the spectral morphology 
and colour of the new T dwarfs and how these could relate to their physical 
properties, and Section 6 presents updated constraints on the size of the 
LAS T dwarf population, compared to theoretical predictions. Section 7 
discusses some future work and gives our conclusions.

\section{Identifying T dwarf candidates}
In this section we describe our photometric search for T dwarfs and cooler 
700--400K objects (potential Y dwarfs). We will focus this description on 
our searches of DR1 and DR2, since our follow-up of candidates from these 
data releases has a reasonably well constrained level of completion, and 
issues such as contamination amongst the candidates will be addressed more 
usefully. However, the process of T dwarf identification is ongoing with 
DR3, using these same techniques. Note also that some DR1 results have 
previously been published (Kendall et al. 2007; Lodieu et al. 2007; Warren 
et al. 2007), but for clarity we here describe our search of DR1 and DR2 as 
a whole.

We base our search on the current knowledge of $izJHK$ properties of 
known objects previously discovered in 2MASS and SDSS, as well as on colour 
trends suggested by the latest theoretical atmosphere models for $T_{\rm eff}$ 
ranges below those that are well probed by previous surveys. We also make 
use of LAS magnitudes measured using the new $Y$ filter (0.97--1.07$\mu$m), 
which was specifically designed and installed in WFCAM to ease the selection 
and separation of high-redshift quasars and cool brown dwarfs (Warren \& 
Hewett 2002).

2MASS T dwarfs have neutral to blue near infrared colours with decreasing 
$T_{\rm eff}$ (Burgasser et al. 2002), and a fairly uniform $Y-J\sim$1 (from 
synthesized colours; Hewett et al. 2006). Late T dwarfs are clearly well 
separated from redder L dwarfs by their $J-H$ colour (although there is 
some $J-H$ colour overlap for early T dwarfs), and from earlier objects 
by, in general, both $J-H$ and $Y-J$ colour (see Figure 1). The model 
predictions suggest that the near infrared colours will remain blue ($J-H<0.0, 
J-K<0.0$) for 400-700K. However, while the models uniformly predict that the 
400-700K dwarf colour sequence should form an extension of the known T dwarfs, 
they differ somewhat on their predicted $Y-J$ colour trends. The cloud-free 
Cond models (Allard et al. 2001; Baraffe et al. 2003), the more recent Settl 
models (Allard et al., in preparation) and the Marley et al. (2002) models 
all suggest a blue $Y-J$ colour trend, while the Burrows et al. (2003) and 
Tsuji, Nakajima \& Yanagisawa (2004) models suggest a red $Y-J$ colour 
trend for this $T_{\rm eff}$ regime.

In addition, known T dwarfs from SDSS have very red ($i-z\ge$2.2) optical 
colours, and extremely red optical-to-infrared colours ($z-J\ge$2.5; note 
that here and hereafter $i$ and $z$ are AB magnitudes whereas $J$ is on the 
Vega system), the latter being a particularly good indicator of $T_{\rm eff}$ 
for early-mid T dwarfs (Knapp et al. 2004). The model predictions suggest that 
these extreme optical and optical-to-infrared colours should continue to be strong 
indicators of low-$T_{\rm eff}$ in the 400-700K regime. A combination of $YJHK$ 
photometry from the LAS and deep optical constraints from SDSS thus offer an 
extremely powerful tool to identify samples of mid-late T dwarfs and even cooler 
objects to photometric depths several magnitudes deeper than previous wide-field 
surveys.

\subsection{Initial sample selection}
Our search methodology was to mine (via the WFCAM Science Archive; Hambly et al. 
2007) the LAS for objects with $YJHK$ photometry consistent with known T dwarfs 
or with the expected colours of 400-700K objects using the models as a guide, 
and then cross-match with SDSS (where possible) to obtain optical-to-infrared 
constraints. Here we describe three sets of search criteria that probe to different 
depths in DR1 and DR2, requiring coverage in all four $YJHK$ bands, but relying 
on source detection in different numbers of bands.

Our first search method required $YJHK$ detection with $Y-J\ge$0.8 and 
$J-H\le$0.4 (to separate T dwarfs from L dwarfs). This method is limited by 
the $K-$band for mid-late T dwarfs, does not probe the full range of $Y-J$ 
colour for potential 400-700K objects, and is thus better at finding brighter 
early T dwarfs which can have redder $H-K$ colour. We also cross-matched 
objects with SDSS, requiring a detection with $i-z\ge$2.2 and $z-J\ge$2.5. 
While we also performed searches requiring optical non-detection, we postpone 
analysis of these candidates to a future publication, and here present the one 
new object identified with a SDSS detection; ULAS J115759.04+092200.7 (see 
Table 1).

Our second search method required $YJH$ LAS detection and a $K$ non-detection, 
and probed the colour space shown in Figure 1, including objects with 
$J-H\le$0.1 and $Y-J\ge$0.5. This search is $H$-band limited for objects 
with blue $J-H$, but fully probes the range of 400-700K colour space suggested 
by the models. Candidates were cross-matched with SDSS and rejected if found 
to be optical detections with either $i-z<$2.2 or $z-J<$2.5, or found to have 
no SDSS coverage. This search produced a sample of 33 candidates.

Our third search method required $YJ$ LAS detection and $HK$ non-detection, 
with $Y-J\ge$0.5. For objects with blue $J-H$ this search is $J$-band or 
$Y$-band limited for $Y-J\le$0.8 and $Y-J\ge$0.8 respectively, and probes 
to a greater depth than method 2. SDSS cross-matching constraints were 
imposed as before, and resulted in a sample of 45 candidates.

In addition we used our second search method to search the LAS sky outside 
the SDSS DR6 footprint, substituting our SDSS optical constraints for 
shallower Schmidt plate $I$-band coverage from the SuperCOSMOS sky survey. 
This search identified 24 candidates.

Figure 1 shows the $J-H$, $Y-J$ two-colour diagram and the $J$ against $J-H$ 
colour magnitude diagram for the T dwarf and $T_{\rm eff}$=700--400K candidates 
from DR1 and DR2. Candidates are either crosses (where $Y-$, $J-$ and $H-$band 
detections were available) or arrows (where $H$ was a non detection). A sample of L 
dwarf candidates (plus signs; LAS $Y-J>$0.9, $J-H>$0.5, $J-K>$1.2, and SDSS $z<$20.8) 
is also shown for comparison (some spectroscopically confirmed L dwarfs from this 
sample will be reported in a separate paper), as is a sample of typical brighter 
sources from the LAS (points). These brighter sources were selected from a one 
degree radius area of DR2 sky at high Galactic latitude, and have YJHK magnitude 
uncertainties less than 0.03 magnitudes and a Sloan counterpart. The L dwarf 
candidates are being followed up, but we will not discuss them further here, 
showing them purely for comparison. Our $YJH$ selection box ($Y-J>$0.5, $J-H<0.1$) 
is indicated with a dashed line, and contains over-plotted boxes (dotted and 
dot-dashed lines) that illustrate the range of model predictions for 
$T_{\rm eff}$=700--400K (see figure caption).

\begin{figure*}
  \includegraphics[width=15cm]{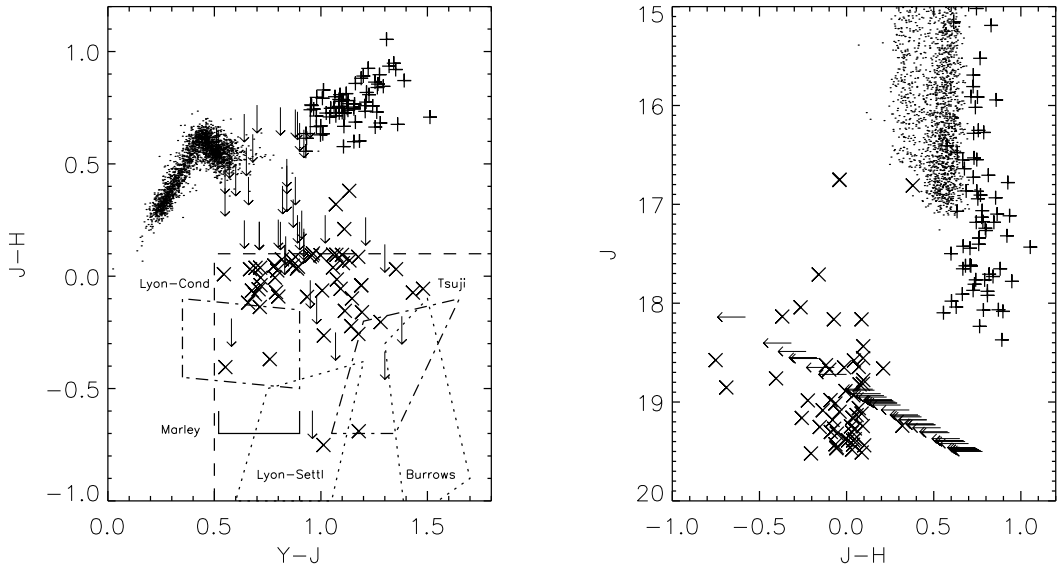}
  \caption{A $J-H$ against $Y-J$ two-colour diagram and a $J$ against $J-H$ colour 
magnitude diagram showing candidate late T and Y dwarfs from DR1 and DR2. $YJH(K)$ 
candidates are shown as crosses, and $YJ$ candidates as upper limit arrows. An 
illustrative L dwarf candidate sample is shown by plus signs. Our $YJH$ candidate 
selection box is shown with a dotted line, and contains theoretical model predicted 
colours for $T_{\rm eff}$=400-700K dwarfs from: the cloud-free Lyon-Cond models 
(Allard et al. 2001; Baraffe et al. 2003); the more recent Lyon-Settl models 
(Allard et al., in preparation); the AMES models (e.g. Marley et al. 2002; 
Saumon et al. 2003); the Tucson models (e.g. Sharp 2000; Burrows Sudarsky \& 
Hubeny 2006); and the Tsuji models (Tsuji Nakajima \& Yanagisawa 2004).}
\end{figure*}

Candidates from our DR1-3 searches that have been confirmed as T dwarfs via 
spectroscopy (see Section 4) are shown in Table 1, which summarises the database 
release in which they were first found, as well as the bands that were used to 
make the searches. All of our new T dwarfs are in SDSS DR6 sky, although two 
previously confirmed candidates (Lodieu et al. 2007) are outside the SDSS DR6 
footprint. Figure 2 presents finding charts for the 15 new T dwarfs.

\begin{table*}
\centering
\caption{Newly discovered UKIDSS T dwarfs. The table lists the coordinates, the 
UKIDSS LAS database release in which objects were found, and the bands that were 
used to make the searches in which the objects were identified. All objects have 
SDSS coverage, although only one is detected in SDSS. The table contains the newly 
confirmed T dwarfs as well as (at the bottom) three that we rule out with spectroscopy.}
\begin{tabular}{lllc}
\hline
Name$^a$                   & Database & NIR band   & SDSS band  \\
                           &          & detections & detections \\
\hline
ULAS J$022200.43-002410.5$ & DR1      & $YJH$      & -          \\
ULAS J$082327.46+002424.4$ & DR3      & $YJHK$     & -          \\
ULAS J$083756.19-004156.0$ & DR1      & $YJH$      & -          \\
ULAS J$085910.69+101017.1$ & DR3      & $YJHK$     & -          \\
ULAS J$093829.28-001112.6$ & DR3      & $YJH$      & -          \\
ULAS J$093951.04+001653.6$ & DR3      & $YJH$      & -          \\
ULAS J$095829.86-003932.0$ & DR1      & $YJ$       & -          \\
ULAS J$115038.79+094942.9$ & DR2      & $YJH$      & -          \\
ULAS J$115759.04+092200.7$ & DR2      & $YJHK$     & $i, z$     \\
ULAS J$130303.54+001627.7$ & DR1      & $YJ$       & -          \\
ULAS J$131508.42+082627.4$ & DR2      & $YJH$      & -          \\
ULAS J$150135.33+082215.2$ & DR2      & $YJ$       & -          \\
ULAS J$150547.89+070316.6$ & DR2      & $YJ$       & -          \\
ULAS J$154427.34+081926.6$ & DR2      & $YJ$       & -          \\
ULAS J$154701.84+005320.3$ & DR2      & $YJH$      & -          \\
\hline
ULAS J$003925.90+000257.7$ & DR1      & $YJH$      & -          \\
ULAS J$130150.35+002314.8^b$ & EDR    & $YJH$      & -          \\
ULAS J$224355.03-004618.5$ & DR1      & $YJH$      & -          \\
\hline
\multicolumn{4}{l}{
$^a$Here the source names include full coordinates. In subsequent tables/figures the}\\
\multicolumn{4}{l}{source names are truncated.} \\
\multicolumn{4}{l}{$^b$First reported as a T dwarf candidate by Kendall et al. (2007).}
\end{tabular}
\end{table*}

\begin{figure*}
\vspace{22.0cm}
  \caption{Finding charts for the 15 new T dwarfs discovered in the UKIDSS 
  LAS and presented in this paper. Charts are $J-$band images of generally two 
  arcmin on a side (unless constrained by the edge of a UKIDSS frame) with 
  north up and east left. Indicated spectral types are infrared types derived 
  here.}
\end{figure*}

\section{Follow-up photometry}

Photometric contamination amongst our candidates has several causes. 
In general, photometry near the survey limits will have significant 
uncertainties due to low signal-to-noise ratio and/or survey image defects 
(see Dye et al. 2006), and in certain circumstances normal stars can 
have their colours scattered to the blue. Scattered late M dwarfs with 
$J>$18 will not always be ruled out via SDSS non detection if their 
$i-J=$3-4 (the 5-$\sigma$ SDSS $i$-band limit is $\sim$22). Also, if 
LAS images in different bands are measured on different nights, variable 
sources can yield extremely blue survey colours, and fast moving solar 
system objects can appear as detections at $Y$ and $J$ while being 
undetected (at that position) in $H$ and $K$ (e.g. LAS $YJ$ observations 
are sometimes taken on different nights to the $HK$ observations of the 
same region). Galaxies can also have unusual colours that could contaminate 
our selections.

We used deeper follow-up $J$- and $H$-band photometry to identify such 
contamination, re-measuring the $J-H$ colour at a higher signal-to-noise 
ratio and identifying any extended sources. We also obtained deeper $Y$-, $K$- 
and $z$-band photometry of a number of our best candidates to provide 
a more complete range of colours with which to assess candidate properties. 
We used a variety of facilities to obtain our infrared and optical 
follow-up photometry, and we summarise these measurements for confirmed 
T dwarfs in Table 2. In general this table presents photometry for the 
newly discovered T dwarfs, but also includes additional $z$-band 
measurements for two previously reported LAS T dwarfs from Lodieu 
et al. (2007).

\begin{table*}
\centering
\caption{Follow-up near infrared and optical photometric observations}
\begin{tabular}{lllll}
\hline
Name              & Instrument  & Filter& UT date  & t$_{int}$ \\
                  & \& telescope&       &          & (m=micro-steps, j=jitters, r=repeats)$^a$ \\
\hline
ULAS J$0222-0024$ & UFTI/UKIRT  & $J$ &  2007 Jul 24 &  300s (j=5, t$_{exp}$=60s) \\
                  &             & $H$ &              & 1800s (j=5, r=6, t$_{exp}$=60s) \\
                  &             & $K$ & 2007 Jul 28 & 1590s (j=5, r=6, t$_{exp}$=53s) \\
                  &             & $Y$ &              &  600s (j=5, t$_{exp}$=120s) \\
                  & EMMI/NTT    & $z_{EMMI}$ & 2007 Nov 17 & 2400s (r=4, t$_{exp}$=600s) \\
ULAS J$0823+0024$ & EMMI/NTT    & $z_{EMMI}$ & 2007 Nov 17 &  900s (r=2, t$_{exp}$=450s) \\
ULAS J$0837-0041$ & WFCAM/UKIRT & $J$ &   2007 May 9 &   320s (m=4, j=2, r=4, t$_{exp}$=10s) \\
                  &             & $H$ &              &   800s (m=4, j=5, r=4, t$_{exp}$=10s) \\
                  &             & $z_{EMMI}$ & 2008 Jan 30 & 2400s (r=4, t$_{exp}$=600s) \\
ULAS J$0859+1010$ & LIRIS/WHT   & $H$ &  2007 Dec 16 &   504s (j=9, r=14, t$_{exp}$=4s) \\
                  & LIRIS/WHT   & $CH_4l$ & 2007 Dec 16 & 450s (j=9, r=5, t$_{exp}$=10s) \\
                  & EMMI/NTT    & $z_{EMMI}$ & 2007 Nov 17 & 600s (r=1, t$_{exp}$=600s) \\
ULAS J$0938-0011$ & EMMI/NTT    & $z_{EMMI}$ & 2008 Jan 29 & 1800s (r=3, t$_{exp}$=600s) \\
ULAS J$0939+0016$ & EMMI/NTT    & $z_{EMMI}$ & 2008 Jan 31 & 900s (r=2, t$_{exp}$=450s) \\
ULAS J$0958-0039$ & LIRIS/WHT   & $J$ &   2007 Mar 1 &  400s (j=5,r=2, t$_{exp}$=40s) \\
                  &             & $H$ &              &  600s (j=5,r=6, t$_{exp}$=20s) \\
                  &             & $K_s$ &          &  800s (j=5,r=8, t$_{exp}$=20s) \\
                  &             & $z_{EMMI}$ & 2008 Jan 30 & 2400s (r=4, t$_{exp}$=600s) \\
ULAS J$1150+0949$ & LIRIS/WHT   & $J$ &   2007 Mar 2 &  200s (j=5, t$_{exp}$=40s) \\
                  &             & $H$ &              & 1200s (j=5,r=12, t$_{exp}$=20s) \\
                  &             & $K_s$ &          & 1800s (j=5,r=19, t$_{exp}$=20s) \\
                  &             & $Y$ &              & 1200s (j=5,r=6, t$_{exp}$=40s) \\
                  &             & $z_{EMMI}$ & 2008 Jan 30 & 2400s (r=4, t$_{exp}$=600s) \\
ULAS J$1157+0922^c$ &           &                  & \\
ULAS J$1303+0016$ & LIRIS/WHT   & $J$ &   2007 Mar 3 &  200s (j=5, t$_{exp}$=40s) \\
                  &             & $H$ &              & 1200s (j=5,r=12, t$_{exp}$=20s) \\
                  &             & $K_s$ &          & 1800s (j=5,r=18, t$_{exp}$=20s) \\
ULAS J$1315+0826$ & LIRIS/WHT   & $J$ &   2007 Mar 2 &  200s (j=5, t$_{exp}$=40s) \\
                  &             & $H$ &              &  400s (j=5,r=4, t$_{exp}$=20s) \\
                  &             & $K_s$ &          & 1000s (j=5,r=10, t$_{exp}$=20s) \\
                  &             & $Y$ &              & 1000s (j=5,r=5, t$_{exp}$=40s) \\
                  &             & $z_{EMMI}$ & 2008 Jan 31 & 2400s (r=4, t$_{exp}$=600s) \\
ULAS J$1501+0822$ & LIRIS/WHT   & $J$ &   2007 Mar 2 &  200s (j=5, t$_{exp}$=40s) \\
                  &             & $H$ &              &  400s (j=5,r=4, t$_{exp}$=20s) \\
                  &             & $K_s$ &          &  800s (j=5,r=8, t$_{exp}$=20s) \\
ULAS J$1505+0703$ & UFTI/UKIRT  & $J$ &  2007 Jul 23 &  540s (j=9, t$_{exp}$=60s) \\
                  &             & $H$ &              & 3240s (j=9, r=6, t$_{exp}$=60s) \\
                  &             & $K$  &         & 3240s (j=9, r=6, t$_{exp}$=60s) \\
                  &             & $Y$ &              & 2160s (j=5, r=2, t$_{exp}$=216s) \\
ULAS J$1544+0819$ & UFTI/UKIRT  & $J$ &  2007 Jul 15 &  300s (j=5, t$_{exp}$=60s) \\
                  &             & $H$ &              & 1800s (j=5, r=6, t$_{exp}$=60s) \\
                  &             & $K$ &          & 1800s (j=5, r=6, t$_{exp}$=60s) \\
                  &             & $Y$ &              & 1080s (j=5, t$_{exp}$=216s) \\
ULAS J$1547+0053$ & LIRIS/WHT   & $J$ &   2007 Mar 2 &  200s (j=5, t$_{exp}$=40s) \\
                  &             & $H$ &              &  600s (j=5,r=6, t$_{exp}$=20s) \\
                  &             & $K_s$ &          &  800s (j=5,r=8, t$_{exp}$=20s) \\
                  &             & $Y$ &              &  800s (j=5,r=4, t$_{exp}$=40s) \\
\hline
ULAS J$0024+0022^b$ & EMMI/NTT    & $z_{EMMI}$ & 2007 Nov 16 & 2400s (r=4, t$_{exp}$=600s) \\
ULAS J$0203-0102^b$ & EMMI/NTT    & $z_{EMMI}$ & 2007 Nov 17 & 1800s (r=3, t$_{exp}$=600s) \\
\hline
\multicolumn{5}{l}{
$^a$Photometric observations consist of a combination of microsteps, jitters (or dithers), 
and repeats, of a single} \\
\multicolumn{5}{l}{
integration time (t$_{exp}$) that combine to give the total integration time (t$_{int}$).}\\
\multicolumn{5}{|l|}{A microstep is an offset of a particular number (e.g. N+1/2) of pixels 
with respect to the current position.}\\
\multicolumn{5}{|l|}{$^b$T dwarfs from Lodieu et al. (2007).}\\
\multicolumn{5}{|l|}{$^c$No photometric follow-up was made of this T dwarf.}
\end{tabular}
\end{table*}

\subsection{Near infrared photometry}
Near infrared follow-up photometry was obtained with three facilities. We used WFCAM 
and the UKIRT Fast-Track Imager (UFTI; Roche et al. 2003) on UKIRT, both of which employ 
the Mauna Kea Observatories (MKO) $J$, $H$ and $K$ filters (Tokunaga, Simons \& Vacca 
2002), as well as the UKIDSS $Y$ filter (Hewett et al. 2006). We also used the Long-slit 
Infrared Imaging Spectrograph (LIRIS; Manchado et al. (1998)) instrument on the William 
Herschel Telescope on La Palma in the Canaries, which uses MKO $J$ and $H$ filters, a $K_s$ 
filter, and a $z_{Liris}$ filter that is quite similar (although slightly narrower band; 
1.00--1.07$\mu$m) to WFCAM $Y$. It also has a narrow band methane filter 
($CH_4l$; 1.64--1.74$\mu$m) that we used to measure one of the new T dwarfs presented here.

Observations comprised a series of jittered (by a few arcsec) image sets, as well 
as (for WFCAM observations) subsets of micro-stepped images that improve the pixel 
sampling during processing. Individual exposure times ranged from 10s to 216s, and 
a summary of the different combinations of micro-step, jitter, repeat, individual 
exposure time and total exposure times, is given in Table 2. All observations were 
made during photometric conditions with good to reasonable seeing (0.6-0.8 arcsec). 
The data was dark-subtracted, flat fielded and mosaiced using standard ORAC-DR routines 
for the UFTI data, LIRIS-DR routines for the LIRIS data, and the Cambridge Astronomical 
Survey Unit's Vista data-flow system (Irwin et al. 2004) for the WFCAM observations. 
Photometry was then performed using typical aperture sizes of $\sim$2-arcsec diameter.

The UFTI observations were calibrated using UKIRT Faint Standards, so the photometric 
system was thus identical to the WFCAM system (Leggett et al. 2006). The WFCAM magnitudes 
were obtained via flat-file access from the WFCAM Science Archive. The LIRIS $J$-, $H$- 
and $K_s$-band observations were calibrated using 2MASS stars as secondary standards where 
$J$ and $H$ magnitudes were converted onto the MKO system via transforms from Warren et al. 
(2007c). 2MASS uses the same $K_s$ filter as LIRIS, so no 2MASS standard $K$ conversion was 
necessary when determining LIRIS $K_s$ magnitudes for our candidates. LIRIS CH$_4l$ photometry 
was calibrated by assuming that the average $H-CH_4l$ colours of 2MASS secondary calibrators 
was zero (see also Kendall et al. 2007). The LIRIS $z$ observations were calibrated using 
observations of A0 stars at similar airmass to our target observations, where 
$(z-J)_{A0}$=0 was assumed.

In order for the T dwarf photometry to be in the same photometric system, we converted 
our LIRIS $z$ and $K_s$ measurements to MKO $Y$ and $K$. Transforms were determined 
via synthetic colours derived from measured T dwarf spectra multiplied by the appropriate 
atmospheric and filter transmission profiles, following the methods of Hewett et al. 
(2006). We derived $Y-z_{Liris}$ and $K-K_s$(LIRIS) colours for the standard T dwarfs 
from B06 as well as for the T dwarfs used in Hewett et al. (2006), excepting objects 
that have been shown to be unresolved binaries (Liu et al. 2006; Burgasser et al. 2006a). 
The $Y-z_{Liris}$ colours show no trend with spectral type, and a conversion of 
$Y$=$z_{Liris}$+(0.015$\pm$0.011) was determined. The $K-K_s$(WHT) colours display 
a tightly defined sequence monotonically increasing with spectral type such that 
$K$=$K_s$(WHT)+(0.032$\times$ST)-0.01, with a scatter of 0.01 magnitudes (where ST=1, 
2, 3... for T1, T2, T3...).

\subsection{Optical photometry}

Optical $z-$band follow-up photometry was obtained with the European Southern Observatory 
(ESO) Multi-Mode Instrument (EMMI) on the New Technology Telescope on La Silla, Chile. 
A Bessel $z-$band filter was used (EMMI\#611; $z_{EMMI}$) which has a short wavelength 
cut-off of 825$\mu$m and is an open-ended filter, where the CCD sensitivity provides 
the longer wavelength cut-off. The conditions were clear, with seeing of $\sim$1 arcsec. 
The EMMI data were reduced using standard IRAF routines including a bias and flat-field 
correction, as well as the removal of fringing effects. Multiple images of the same sky 
areas were then aligned and co-added. Aperture photometry was measured using $\sim$1.5 
arcsec apertures, and we calibrated this photometry using Sloan sources (York et al. 2000) 
contained in the images as secondary standards. This method allowed us to derive image 
zero points with a typical accuracy of $\pm$0.05 magnitudes. We transformed the Sloan AB 
photometry of our secondary standards into the EMMI system using the transformation of 
Warren et al. (2007a). We then derived $z_{EMMI}$ (AB) photometry for the targets, and 
then transformed these into the Sloan AB magnitudes assuming that $z$(AB)=$z_{EMMI}$(AB)+0.2 
(Warren et al. 2007a). Our optical $z-$band photometry is thus all in the Sloan AB system.

\subsection{Photometric results}

We have been able to reject 61 of our 103 DR1 and DR2 candidates via follow-up photometry, 
since the $J-H$ colours were found to be $\ge$0.4, and the candidates thus likely M or L 
dwarfs. A total of 23 DR1 and DR2 T dwarfs have been confirmed amongst the remaining candidates, 
including 11 of the 15 new T dwarfs reported here (four are from DR3), and 12 previously 
reported: one by Geballe et al. (2002), one by Kendall et al. (2007), one by Warren et al. 
(2007a), eight by Lodieu et al (2007), and one by Chiu et al. (2007). Near infrared and 
optical photometry for all 15 newly discovered T dwarfs is shown in Table 3, and their 
spectroscopic confirmation is discussed in Section 4. One of the remaining candidates has 
follow-up photometric colours that suggest it is an additional T dwarf (but no spectrum has 
been obtained to date), and 18 candidates still await follow-up. Further discussion of the 
magnitude limited completeness of our follow-up will be presented in Section 6.


\begin{landscape}
\begin{table}
\caption{T dwarf photometry and colours. Unless otherwise indicated, near 
infrared photometry was measured on the MKO system. Uncertainties are indicated 
in brackets as integer multiples of the last decimal place. The last two T dwarfs 
are from Lodieu et al. (2007), and here we present additional $z$-band measurements.}
\begin{tabular}{lccccccccc}
\hline
Name              & $Y$                & $J$                & $H$               & $K$            & $z$        & $Y-J$     & $J-H$      & $H-K$      & $z-J$ \\
\hline
ULAS J$0222-0024$ & 19.87$\pm$0.03     & 18.71$\pm$0.02     & 19.02$\pm$0.02    & 19.18$\pm$0.03 & 22.50$\pm$0.07$^a$ & 1.16$\pm$0.04 & -0.31$\pm$0.03 & -0.16$\pm$0.04 & 3.79$\pm$0.07 \\
ULAS J$0823+0024$ & 19.93$\pm$0.15$^b$ & 18.57$\pm$0.05$^b$ & 18.96$\pm$0.18$^b$ & 18.58$\pm$0.23$^b$ & 22.80$\pm$0.15 & 1.36$\pm$0.16 & -0.39$\pm$0.19 &  0.38$\pm$0.29 & 4.23$\pm$0.16 \\
ULAS J$0837-0041$ & 19.64$\pm$0.03$^b$ & 18.52$\pm$0.09     & 18.60$\pm$0.11    &                & 22.18$\pm$0.06   & 1.12$\pm$0.09  & -0.08$\pm$0.14 &            & 3.66$\pm$0.11 \\
ULAS J$0859+1010^c$ & 19.00$\pm$0.07$^b$ & 17.88$\pm$0.06$^b$ & 18.58$\pm$0.06  & 18.26$\pm$0.15$^b$ & 21.53$\pm$0.05  & 1.12$\pm$0.09  & -0.70$\pm$0.08  &  0.32$\pm$0.16 & 3.65$\pm$0.08 \\
ULAS J$0938-0011$ & 19.83$\pm$0.12$^b$ & 18.53$\pm$0.06$^b$ & 19.00$\pm$0.19$^b$ &               & 22.27$\pm$0.05   & 1.30$\pm$0.13 & -0.47$\pm$0.20 &             & 3.74$\pm$0.08 \\
ULAS J$0939+0016$ & 19.20$\pm$0.07$^b$ & 17.96$\pm$0.03$^b$ & 18.41$\pm$0.11$^b$ &               & 21.74$\pm$0.05   & 1.24$\pm$0.08  & -0.45$\pm$0.11 &            & 3.78$\pm$0.06 \\
ULAS J$0958-0039$ & 19.88$\pm$0.15$^b$ & 18.95$\pm$0.06     & 19.40$\pm$0.10    & 19.68$\pm$0.12$^d$ & 22.85$\pm$0.09   & 0.93$\pm$0.16 & -0.45$\pm$0.12 & -0.28$\pm$0.16 & 3.90$\pm$0.10 \\
ULAS J$1150+0949$ & 19.92$\pm$0.08$^e$ & 18.68$\pm$0.04     & 19.23$\pm$0.06    & 19.06$\pm$0.05$^d$ & 22.44$\pm$0.10   & 1.24$\pm$0.09  & -0.55$\pm$0.07  &  0.17$\pm$0.08  & 3.76$\pm$0.11 \\
ULAS J$1157+0922$ & 17.94$\pm$0.03$^b$ & 16.81$\pm$0.01$^b$ & 16.43$\pm$0.02$^b$ & 16.24$\pm$0.04$^b$ &            & 1.13$\pm$0.03  &  0.38$\pm$0.02  &  0.19$\pm$0.04  &  \\
ULAS J$1303+0016$ & 20.21$\pm$0.17$^b$ & 19.02$\pm$0.03     & 19.49$\pm$0.09    & 20.10$\pm$0.17$^d$ &            & 1.19$\pm$0.17 & -0.47$\pm$0.09  & -0.61$\pm$0.19 &  \\
ULAS J$1315+0826$ & 20.00$\pm$0.08$^e$ & 18.86$\pm$0.04     & 19.50$\pm$0.10    & 19.60$\pm$0.12$^d$ & 22.82$\pm$0.10   & 1.14$\pm$0.09  & -0.64$\pm$0.11 & -0.10$\pm$0.16 & 3.96$\pm$0.11 \\
ULAS J$1501+0822$ & 19.70$\pm$0.15$^b$ & 18.32$\pm$0.02     & 18.30$\pm$0.06    & 18.53$\pm$0.10$^d$ &            & 1.38$\pm$0.15 &  0.02$\pm$0.06  & -0.23$\pm$0.12 &  \\
ULAS J$1505+0703$ & 20.32$\pm$0.02     & 18.96$\pm$0.03     & 19.10$\pm$0.03    & 19.29$\pm$0.03 &            & 1.36$\pm$0.04  & -0.14$\pm$0.04  & -0.19$\pm$0.04  &  \\
ULAS J$1544+0819$ & 19.80$\pm$0.05     & 18.53$\pm$0.03     & 18.49$\pm$0.03    & 18.73$\pm$0.03 &            & 1.27$\pm$0.06  &  0.04$\pm$0.03  & -0.24$\pm$0.04  &  \\
ULAS J$1547+0053$ & 19.37$\pm$0.06$^e$ & 18.32$\pm$0.03     & 18.45$\pm$0.07    & 18.21$\pm$0.10$^d$ &            & 1.05$\pm$0.07  & -0.13$\pm$0.08  &  0.24$\pm$0.12 &  \\
\hline
ULAS J$0024+0022^f$ &                  &                    &                   &                & 22.77$\pm$0.10$^a$ &       &            &            & 4.61$\pm$0.10 \\
ULAS J$0203-0102^f$ &                  &                    &                   &                & 22.11$\pm$0.06$^a$ &       &            &            & 4.07$\pm$0.06 \\
\hline
\multicolumn{10}{|l|}{$^az_{EMMI}$ converted into z$_{SDSS}$.}\\
\multicolumn{10}{|l|}{$^b$Photometry from the UKIDSS LAS.}\\
\multicolumn{10}{|l|}{$^c$Also has a measured $H-CH_4l$ colour of -1.05$\pm$0.12, typical of mid-late T dwarfs.}\\
\multicolumn{10}{|l|}{$^dK_s$(Liris) converted into K(MKO).}\\
\multicolumn{10}{|l|}{$^ez_{Liris}$ converted into Y(MKO).}\\
\multicolumn{10}{|l|}{$^f$T dwarfs from Lodieu et al. (2007).}
\end{tabular}
\end{table}
\end{landscape}

\section{Spectroscopic observations}
Spectroscopic follow-up observations were made at five different facilities, providing 
a variety of near-infrared spectral ranges and sensitivities. Table 4 summarises which 
facilities were used to observe the newly confirmed T dwarfs in this work (as well as 
three candidates that were ruled out by spectroscopy), and gives details on the wavelength 
range covered, the date of the observations and the total exposure times used. A description 
of these facilities and the reduction methods we used is given in the next sections.

\begin{table*}
\centering
\caption{Spectroscopic observations. The table contains the new confirmed T dwarfs, as well 
as (at the bottom) three that we rule out with spectroscopy.}
\begin{tabular}{lllrr}
\hline
Name & Instrument  & Wavelength & UT date & t$_{int}$ \\
     & \& telescope& range      &         &           \\
\hline
ULAS J$0222-0024$ & NIRI/Gemini-N   & 1.05-1.41$\mu$m &  2 Sep 2007 &  960s (4$\times$240s) \\
ULAS J$0823+0024$ & NIRI/Gemini-N   & 1.05-1.41$\mu$m &  2 Sep 2007 &  960s (4$\times$240s) \\
ULAS J$0837-0041$ & NIRI/Gemini-N   & 1.05-1.41$\mu$m &  5 Oct 2007 &  960s (4$\times$240s) \\
ULAS J$0859+1010$ & NIRI/Gemini-N   & 1.05-1.41$\mu$m &  2 Sep 2007 &  960s (4$\times$240s) \\
ULAS J$0938-0011$ & NIRI/Gemini-N   & 1.05-1.41$\mu$m &  2 Sep 2007 &  960s (4$\times$240s) \\
ULAS J$0939+0016$ & NIRI/Gemini-N   & 1.05-1.41$\mu$m &  2 Sep 2007 &  960s (4$\times$240s) \\
ULAS J$0958-0039$ & GNIRS/Gemini-S  &   0.9-2.5$\mu$m &  4 Apr 2007 &  960s (4$\times$240s) \\
ULAS J$1150+0949$ & GNIRS/Gemini-S  &   0.9-2.5$\mu$m & 10 Mar 2007 &  960s (4$\times$240s) \\
ULAS J$1157+0922$ & NICS/TNG        &   0.9-2.5$\mu$m &  9 Apr 2007 & 1200s (4$\times$300s) \\
                  & UIST/UKIRT      &   1.4-2.5$\mu$m & 27 Jun 2007 & 5760s (24$\times$240s) \\
ULAS J$1303+0016$ & GNIRS/Gemini-S  &   0.9-2.5$\mu$m & 10 Mar 2007 &  960s (4$\times$240s) \\
ULAS J$1315+0826$ & GNIRS/Gemini-S  &   0.9-2.5$\mu$m &  4 Apr 2007 &  960s (4$\times$240s) \\
ULAS J$1501+0822$ & IRCS/Subaru     &     1-1.6$\mu$m &  1 Jul 2007 & 2400s (8$\times$300s) \\
ULAS J$1505+0703$ & NIRI/Gemini-N   & 1.05-1.41$\mu$m & 22 Aug 2007 & 1200s (4$\times$300s) \\
ULAS J$1544+0819$ & NIRI/Gemini-N   & 1.05-1.41$\mu$m &  4 Sep 2007 &  960s (4$\times$240s) \\
ULAS J$1547+0053$ & IRCS/Subaru     &     1-1.6$\mu$m &  1 Jul 2007 & 1920s (8$\times$240s) \\
\hline
ULAS J$0039+0002$ & IRCS/Subaru     &     1-1.6$\mu$m &  1 Jul 2007 & 1920s (8$\times$240s) \\
ULAS J$1301+0023$ & IRCS/Subaru     &     1-1.6$\mu$m &  1 Jul 2007 & 1920s (8$\times$240s) \\
ULAS J$2243-0046$ & IRCS/Subaru     &     1-1.6$\mu$m &  1 Jul 2007 & 1920s (8$\times$240s) \\
\hline
\end{tabular}
\end{table*}

\subsection{Gemini/GNIRS spectroscopy}
The Gemini Near-Infrared Spectrograph (GNIRS; Elias et al. 2006) on Gemini South 
was used to make quick response observations, through programme GS-2007A-Q-15. 
GNIRS was used in cross-dispersed mode with the 32l~mm$^{-1}$ grism, the 1.0-arcsec 
slit and the short camera, to obtain 0.9--2.5$\mu$m R$\simeq$500 (per resolution 
element) spectra. The targets were nodded three arcsec along the slit in an ABBA pattern 
using individual exposure times of 240s. Calibrations were achieved using lamps in 
the on-telescope calibration unit. A0 and early F stars were observed as spectroscopic 
standards, either directly before or after the target observations, at an airmass 
that closely matched the mid-point airmass of the target, in order to remove the 
effects of telluric absorption. The observing conditions included some patchy cloud, 
seeing from 0.5--1.0 arcsec, and humidity ranging from 10--50 per cent.

Data reduction was initially implemented using tasks in the Gemini GNIRS IRAF 
(Image Reduction and Analysis Facility) package. Files were prepared and corrected 
for offset bias using NSPREPARE and NVNOISE, and order separation achieved with NSCUT. 
Each order was then median stacked at the A and B positions, and a difference image 
obtained using GEMARITH. Flat-field correction was not necessary since variations across 
the 6-arcsec slit are less than 0.1 per cent, and any flat-field variations in the 
dispersion direction will subsequently be removed when dividing by the spectrum of 
a standard. S-distortion correction and wavelength calibration were performed 
interactively using the telluric star spectra and argon arc lamp spectra, with 
NSAPPWAVE, NSSDIST and NSWAVE. Further reduction was carried out using custom 
written IDL (Interactive Data Language) routines. Apertures (1.5-arcsec wide) were 
centred on the spectra at the A and B positions, and the sky residuals were fit (and 
subtracted) using a surface constructed via a series of least-squares linear fits 
across the slit (excluding pixels within the apertures), with one fit for each 
spatial pixel row. Spectra were then extracted by summing within the A and B position 
apertures and combining. The target spectra were flux calibrated on a relative scale 
using the telluric standard spectra (after appropriate interpolation across any 
hydrogen absorption lines) with an assumed black-body function for $T_{\rm eff}$=10000 and 
7000~K for A0 and early F tellurics (e.g. Masana et al. 2006), respectively. The 
spectral orders were then trimmed of their noisiest portions, and the spectra 
normalized to unity at 1.27$\pm0.005\mu$m.

\subsection{Gemini/NIRI spectroscopy}
Gemini's Near Infrared Camera and Spectrometer (NIRI; Hodapp et al. 2003) was used, 
on the Gemini North Telescope on Mauna Kea, Hawaii, through programme GN-2007B-Q-26. 
NIRI was used in the f/6 mode with a 0.75 arcsec slit, and with the $J$ grism (312.6 
l~mm$^{-1}$) and G0209 order sorting filter. This produced R$\sim$460 spectra over 
1.05--1.41$\mu$m. The targets were nodded 10 arcsec along the slit in an ABBA pattern, 
and calibrations were achieved using argon lamp observations (as with GNIRS), and F 
stars as telluric standards. The observing conditions were very similar to those of 
the GNIRS observations.

NIRI data was reduced using tasks in the Gemini NIRI IRAF package. NRESID was used to 
create bad pixel masks from the individual flat-field observations. Combined flat-fields 
were created with NSFLAT, and then bad-pixel removal and flat fielding of the science 
data was achieved using NSREDUCE. Sky removal was then carried out by subtracting 
consecutive AB pairs with GEMARITH, and multiple spectra combined with NSSTACK. We 
also minimised any pattern noise in our data using a custom written Python script, 
and wavelength calibrated with NSWAVE and corrected for any S-distortion with NSSDIST 
and NSTRANSFORM. Spectra were then extracted and calibrated using the same methods 
and software as were used for our GNIRS observations.

\subsection{Subaru/IRCS spectroscopy}
We used the Infrared Camera Spectrograph (IRCS; Kobayashi et al. 2000) on the Subaru 
Telescope on Mauna Kea, Hawaii. IRCS was used with its camera in the 52mas/pixel mode, 
and with the JH grism and a 0.6-arcsec slit. This resulted in R$\sim$100 spectra over 
a wavelength range of  1.0--1.6$\mu$m. The targets were nodded seven arcsec along the 
slit in an ABBA pattern. Calibrations were achieved using an argon lamp, and F2-5 stars 
were observed as telluric standards. The observing conditions were clear with seeing of 
$\sim$0.7 arcsec.

The IRCS data was reduced following the same reduction procedures as were used for our 
NIRI data, but using generic spectral reduction IRAF packages. The only practical 
difference was that S-distortion correction and wavelength calibration were done at 
the same time, using the TRANSFORM package. Arc observations were obtained several times 
throughout the night, and we performed wavelength calibration with the arc closest in time 
to each target observation. We estimate wavelength residuals (by inter-comparing arcs) 
of $\sim$20\AA\, which should not affect significantly our subsequent analysis.

\subsection{UKIRT/UIST spectroscopy}
One of the T dwarfs (ULAS J1157+0922) was observed at UKIRT using the UKIRT Imager 
Spectrometer (UIST; Ramsay Howat et al. 2004). The HK grism was used with the 4-pixel 
slit, giving a resolution R = 550. Individual exposure times on target were 240 s, and 
the target was nodded up and down the slit by 12 arcsec. The instrument calibration lamps 
were used to provide accurate flat-fielding and wavelength calibration. The F5V star 
HD 96218 was observed prior to the target to remove the effects of telluric absorption, 
and to provide an approximate flux calibration. Relative flux calibration was improved 
by scaling the spectra to the $H$ and $K$ magnitudes measured in the LAS, and finally we 
scaled the spectra at 1.6$\mu$m to join onto our broader wavelength TNG spectra (see 
next section).

\subsection{TNG/NICS spectroscopy}
We have also obtained a low-resolution (R$\sim$50) near-infrared (0.9--2.5$\mu$m) 
spectrum of ULAS J1157+0922 with the Near Infrared Camera Spectrometer (NICS; Baffa 
et al. 2001) on the Telescope Nazionale Galileo (TNG). The observations were made in 
service mode by the TNG staff, with seeing in the 1.2--2.0 arcsec range. NICS is a 
multi-purpose instrument equipped with a HgCdTe Hawaii 1024x1024 detector with 0.25 
arcsec per pixel, yielding a 4.2 by 4.2 arcmin field-of-view. We employed the Amici 
mode with a one arcsec slit, yielding a wavelength coverage of 0.9--2.5 $\mu$m. Four 
integrations of 300 sec were taken for the target in an ABBA dither pattern. A and B 
spectra were averaged, differenced, and the resulting image flat-fielded. A one-dimensional 
spectrum was then extracted with the IRAF task APSUM\@. A standard star (HIP48971; A0) 
was observed immediately after the target at a similar airmass, and divided through 
the science target to correct for telluric absorption, before flux calibrating with 
a black-body of the same $T_{\rm eff}$ as the A0 star. Due to the low resolution of 
the Amici mode, virtually all the Ar/Xe arc lines are blended and cannot be easily 
used for standard reduction procedures. For this reason, wavelength calibration was 
performed in a NICS-specific way, using a look-up table (of wavelength against pixel) 
based on the instrument's theoretical dispersion from ray-tracing. This first-pass 
wavelength calibration was then offset to best fit the observed spectra of the calibration 
sources. No attempt was made to correct the very modest level of slit curvature.

\subsection{Spectral classification}

\begin{figure*}
  \includegraphics[width=15cm]{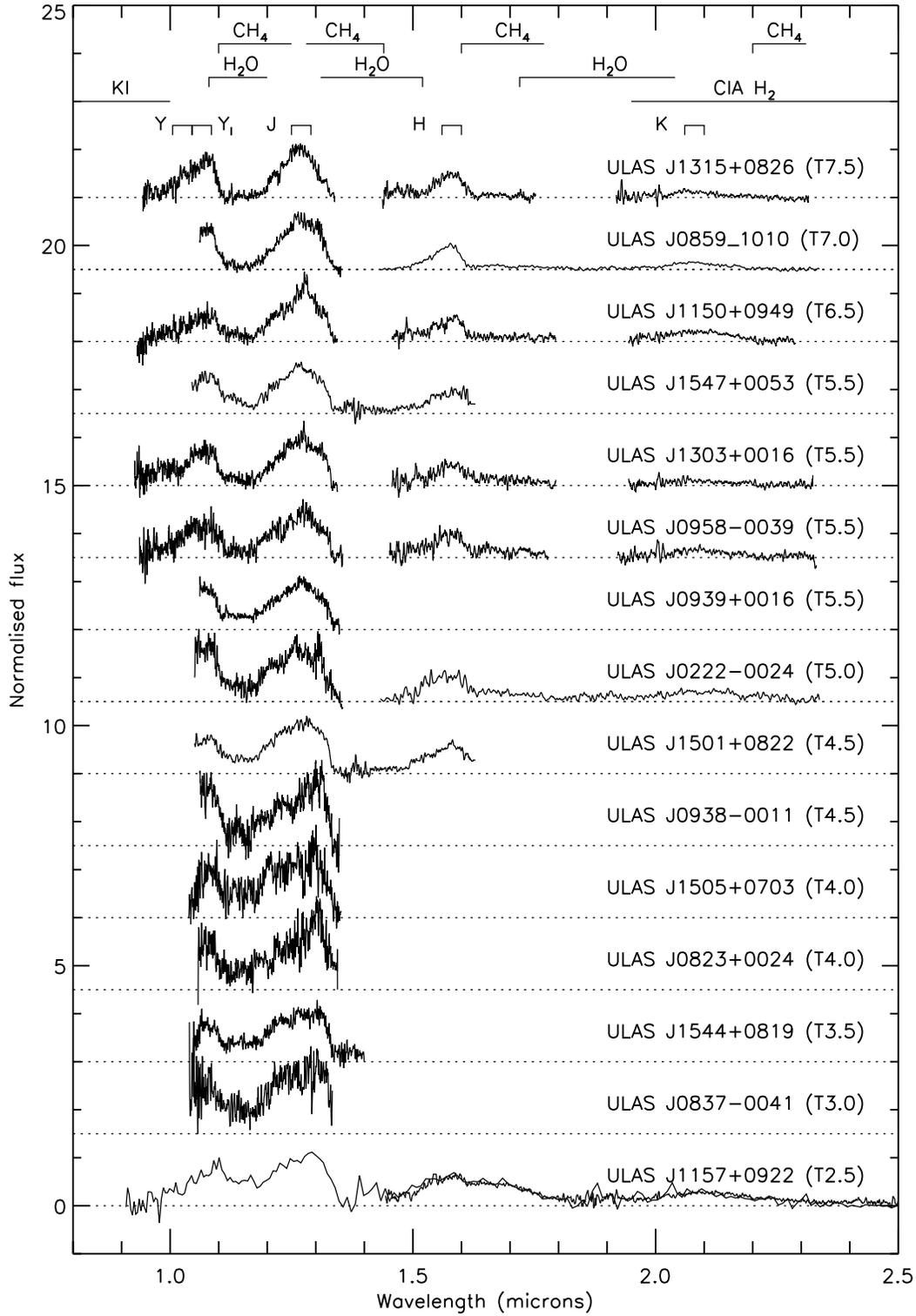}
  \caption{Spectra of the 15 new T dwarfs. The spectra are all normalised to unity 
  at 1.27$\pm0.005\mu$m. Defining features of T dwarf spectra are indicated. Also 
  shown are the $Y$, $Y_l$, $J$, $H$ and $K$ bands which we use to calculate spectral 
  peak ratios (see Section 5).}
\end{figure*}

Figure 3 shows the spectra of the 15 new T dwarfs. 
For each spectrum we calculated all available near-infrared spectral indices from 
B06; H2O-J, CH4-J, H2O-H, CH4-H and CH4-K. For the GNIRS and NICS spectra 
we were able to measure all five ratios. For the IRCS spectra we were able to measure 
H$_2$O-J, CH$_4$-J and H$_2$O-H. The NIRI spectra allowed us to measure the H$_2$O-J 
and CH$_4$-J, and the UIST spectrum covered H$_2$O-H, CH$_4$-H and CH$_4$-K. We also 
compared the data by eye to template spectra of T dwarfs defined by B06 as standard 
T4, T5, T6, T7 and T8 types. The results for the new T dwarfs are 
given in Table 5. Some scatter is present in the spectral indices given in Table 5 
due to low signal-to-noise ratio. Consequently, the direct comparison with templates 
has been given more weight than the spectral indices in the assignment of spectral types, 
and the adopted uncertainty in Table 5 reflects the range in type implied by this comparison. 
Note also that in the case of one object (ULAS J$1150+0949$), the H$_2$O-H spectral ratio 
indicates a significantly earlier spectral type (T3) than all the other ratios for this 
object (T6-7), and we have thus typed it as T6.5p. Table 5 also presents distance estimates 
for the T dwarfs based on their spectral type and $J$-band brightness, using the $M_J$ 
spectral type relation (excluding known and possible binaries) of Liu et al. (2006), and 
allowing for the rms scatter in this relation.

\begin{table*}
\centering
\caption{Spectral types derived from indices and by comparison with T dwarf 
templates (following B06).}
\begin{tabular}{lllllllll}
\hline
Name & H$_2$O-J & CH$_4$-J & H$_2$O-H & CH$_4$-H & CH$_4$-K & Template & Adopted & Distance$^a$ (pc) \\
\hline
ULAS J$0222-0024$ & 0.272 (T5)  & 0.368 (T5.5)&             &
             &                & T5        & T5.0$\pm$0.5 & 55-78 \\
ULAS J$0823+0024$ & 0.390 (T3)  & 0.696 (T1)  &   &
             &                & T4$\pm$1  & T4$\pm$1 & 55-78 \\
ULAS J$0837-0041$ & 0.419 (T3)  & 0.660 (T2)  &             &
             &                & T3        & T3.0$\pm$0.5 & 54-76 \\
ULAS J$0859+1010$ & 0.058 (T8)  & 0.266 (T7)  &   &
             &                & T7     & T7$\pm$0.5 & 26-36 \\
ULAS J$0938-0011$ & 0.190 (T6)  & 0.484 (T4)  &   &
             &                & T4.5   & T4.5$\pm$0.5 & 53-75 \\
ULAS J$0939+0016$ & 0.288 (T5)  & 0.366 (T5)  &   &
             &                & T5-6   & T5.5$\pm$0.5 & 37-52 \\
ULAS J$0958-0039$ & 0.125 (T6.5)& 0.341 (T5.5)& 0.391 (T4)  &
 0.402 (T5)  & 0.130 (T6.5)   & T5.5$\pm$1& T5.5$\pm$0.5 & 58-82 \\
ULAS J$1150+0949$ & 0.087 (T7)  & 0.302 (T6.5)& 0.455 (T3)  &
 0.230 (T6.5)& 0.032 ($\ge$T7)& T6-7      & T6.5p$\pm$0.5 & 42-60 \\
ULAS J$1157+0922$ & 0.428 (T3)  & 0.544 (T3)  & 0.453 (T3)  &
 0.772 (T3)  & 0.557 (T2)     & T2        & T2.5$\pm$0.5 & 24-34 \\
ULAS J$1303+0016$ & 0.105 (T7)  & 0.324 (T6)  & 0.332 (T5.5)&
 0.367 (T5.5)& 0.025 ($\ge$T7)& T5-6      & T5.5$\pm$0.5 & 60-85 \\
ULAS J$1315+0826$ & 0.034 (T8)  & 0.181 (T8)  & 0.227 (T7)  &
 0.121 (T8)  & 0.080 ($>$T7)  & T8        & T7.5$\pm$0.5 & 34-48 \\
ULAS J$1501+0822$ & 0.290 (T5)  & 0.418 (T5)  & 0.411 (T4)  &
             &                & T4-5      & T4.5$\pm$0.5 & 48-68 \\
ULAS J$1505+0703$ & 0.349 (T4)  & 0.572 (T2.5)&             &
             &                & T4$\pm$1  & T4.0$\pm$0.5 & 66-93 \\
ULAS J$1544+0819$ & 0.411 (T3)  & 0.520 (T3.5)&             &
             &                & T3.5      & T3.5$\pm$0.5 & 55-77 \\
ULAS J$1547+0053$ & 0.217 (T5)  & 0.363 (T5)  & 0.331 (T5)  &
             &                & T5-6      & T5.5$\pm$0.5 & 44-61 \\
\hline
\multicolumn{9}{l}{$^a$Assumes single objects.}\\
\end{tabular}
\end{table*}

\begin{figure}
  \includegraphics[width=9cm]{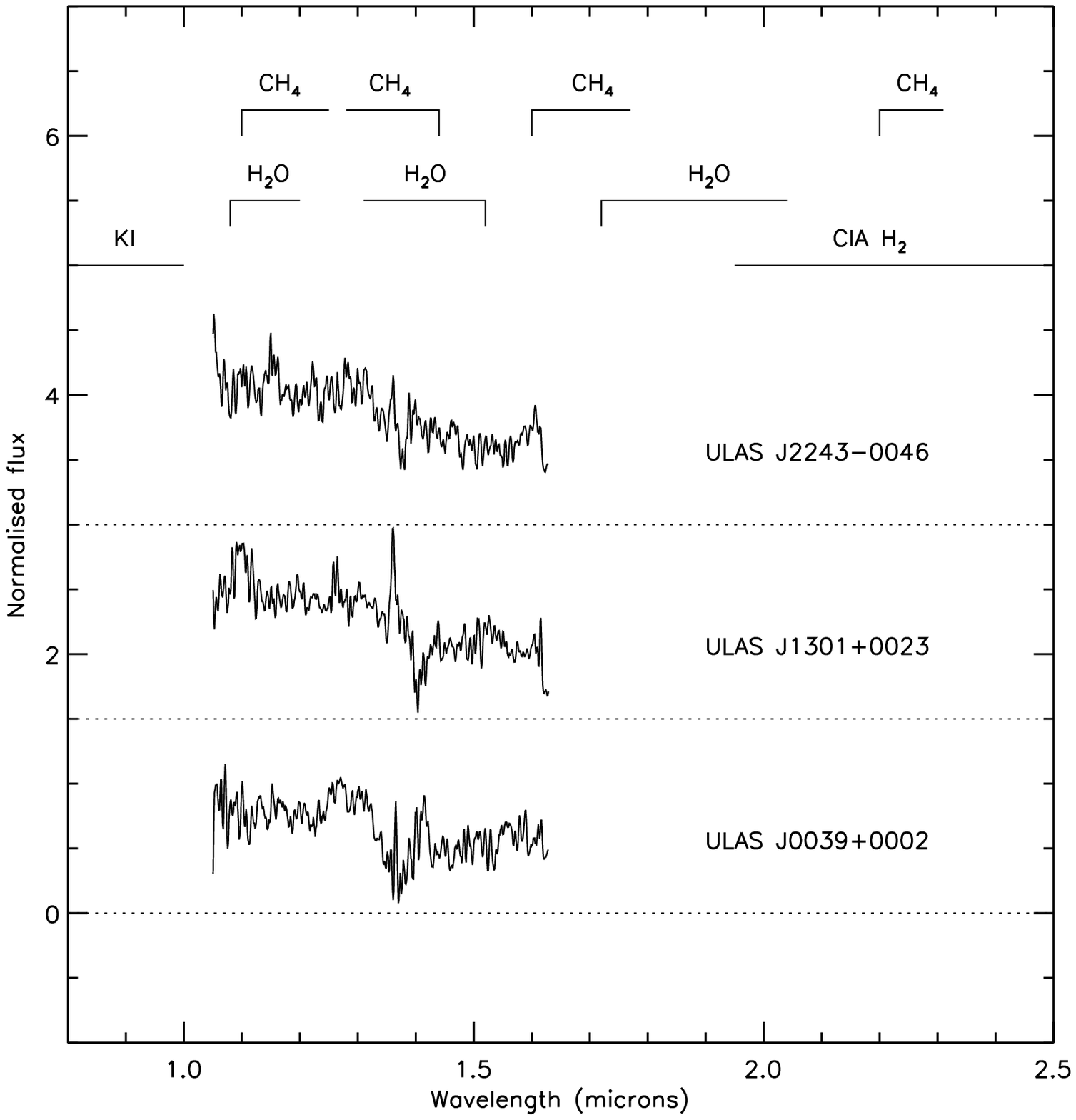}
  \caption{Spectra of the three T dwarf candidates ruled out with spectroscopy. 
The absorption feature that we expect to define T dwarf spectra are indicated, and 
are clearly not present in these spectra.}
\end{figure}

Figure 4 shows the Subaru spectra of the three candidate T dwarfs that we 
rule out with spectroscopy. Clearly none of these objects show the strong 
H$_2$O and CH$_4$ absorption at 1.1--1.2$\mu$m and 1.3--1.5$\mu$m that 
sculpts the $J-$band and $H-$band peaks characteristic of T dwarfs. After 
allowing for some residual telluric noise at $\sim$1.4$\mu$m, the spectrum 
of ULAS J2243-0046 appears to have the form of a blue continuum. The other 
two sources show evidence of H$_2$O absorption at 1.3--1.4$\mu$m, and appear 
to be M dwarfs. The LAS YJ and HK photometry used to select these candidates 
was taken on two separated nights, and variability might explain how some 
M dwarfs can appear to have blue $J-H$ colours. ULAS J1301+0023 also showed 
some evidence for possible methane absorption (at the $\sim$1--2$\sigma$ 
level) via narrow-band CH$_4$s- and CH$_4$l-band photometry (Kendall et al. 
2007). Our spectrum of this object does not cover the full wavelength range 
of these filters (1.53--1.63$\mu$m and 1.64--1.74$\mu$m respectively), so we 
cannot rule out an intrinsically blue methane ($CH_4$s-$CH_4$l) colour. 
However, it may simply be that the relatively low signal-to-noise ratio of 
the narrow-band photometry led directly to the measurement of a blue colour.

\section{T dwarf properties}

We now consider the overall spectral morphology of the T dwarfs using both 
their spectra and their photometric properties. We measured the relative 
brightness of the $Y-$, $J-$, $H-$, and $K-$band spectral peaks by summing flux 
in several bands. In general we used the bands defined by Burgasser, Burrows 
\& Kirkpatrick (2006), to which we also add an additional $Y-$band at a 
slightly longer wavelength (which we refer to as $Y_l$). Our $Y-$, $Y_l$-, 
$J-$, $H-$, and $K-$band spectral peaks cover the 1.005--1.045, 1.045--1.085, 
1.25--1.29, 1.56--1.60, and 2.06--2.10$\mu$m wavelength ranges respectively, 
and are indicated in Figure 5. Table 6 shows the integrated flux peak ratios 
for the 15 new T dwarfs. We were able to measure the $Y_l$/J ratio for all our 
T dwarf spectra, while the short wavelength cut-off in the spectral coverage 
of several of the T dwarfs precludes the measurement of the $Y/J$ ratio.

\begin{table*}
\centering
\caption{Ratios of spectral peaks.}
\begin{tabular}{llllllll}
\hline
Name & SpT & Y/J & Y$_l$/J & H/J & K/J & K/H & Unusual property?\\
\hline
ULAS J$0222-0024$ & T5.0 & -     & 0.997 & -     & -     & -     & Noisy - possibly unusual metallicity \\
ULAS J$0823+0024$ & T4.0 & -     & 0.892 & -     & -     & -     & Normal \\
ULAS J$0837-0041$ & T3.0 & -     & 0.833 & -     & -     & -     & Normal \\
ULAS J$0859+1010$ & T7.0 & -     & 0.814 & -     & -     & -     & Normal \\
ULAS J$0938-0011$ & T4.5 & -     & 1.195 & -     & -     & -     & Noisy - possibly unusual metallicity \\
ULAS J$0939+0016$ & T5.5 & -     & 0.918 & -     & -     & -     & Normal \\
ULAS J$0958-0039$ & T5.5 & 0.488 & 0.708 & 0.521 & 0.182 & 0.349 & Noisy - possibly low $T_{\rm eff}$ \\
                  &      &       &       &       &       &       & or unusual metallicity \\
ULAS J$1150+0949$ & T6.5 & 0.299 & 0.467 & 0.428 & 0.199 & 0.465 & Low-gravity$^a$ \\
ULAS J$1157+0922$ & T2.5 & 0.401 & 0.467 & 0.652 & 0.291 & 0.446 & Normal \\
ULAS J$1303+0016$ & T5.5 & 0.416 & 0.738 & 0.449 & 0.110 & 0.245 & High gravity \\
ULAS J$1315+0826$ & T7.5 & 0.436 & 0.723 & 0.472 & 0.122 & 0.258 & Normal \\
ULAS J$1501+0822$ & T4.5 & -     & 0.658 & 0.526 & -     & -     & Normal \\
ULAS J$1505+0703$ & T4.0 & -     & 0.710 & -     & -     & -     & Normal \\
ULAS J$1544+0819$ & T3.5 & -     & 0.638 & -     & -     & -     & Normal \\
ULAS J$1547+0053$ & T5.5 & -     & 0.739 & 0.437 & -     & -     & High gravity$^b$ \\
\hline
\multicolumn{8}{l}{$^a$Possibly high [M/H] as well.}\\
\multicolumn{8}{l}{$^b$From photometric colour only.}
\end{tabular}
\end{table*}

To assess how the spectral morphology of the T dwarfs may depend on their physical 
properties, we compare our observed properties with the predictions of model atmospheres. 
As a primary comparison we use the BT-Settl models (which combine with structure models 
to give the Lyon-Settl models shown in Figure 1), generated with version 15.3 of the 
general-purpose stellar atmosphere code Phoenix (Hauschildt \& Baron 1999). These models 
use a set-up that currently gives the best BT-Settl fits to observed spectra of M, L and 
T dwarfs, updating the micro-physics used in the GAIA model grid (Kucinskas et al. 2005, 
2006). For a summary of the important input physics of these models, see Warren et al. 
(2007a). Note that Burgasser, Burrows \& Kirkpatrick (2006) and Leggett et al. (2007) 
have also examined these effects in the spectral models of Burrows et al. (2006; Tucson 
models) and Marley et al. (in prep.; AMES models), and while the different model trends 
in spectral morphology are in broad agreement, we highlight any differences between 
these model sets and the BT-Settl models.

\begin{figure*}
  \includegraphics[width=15cm]{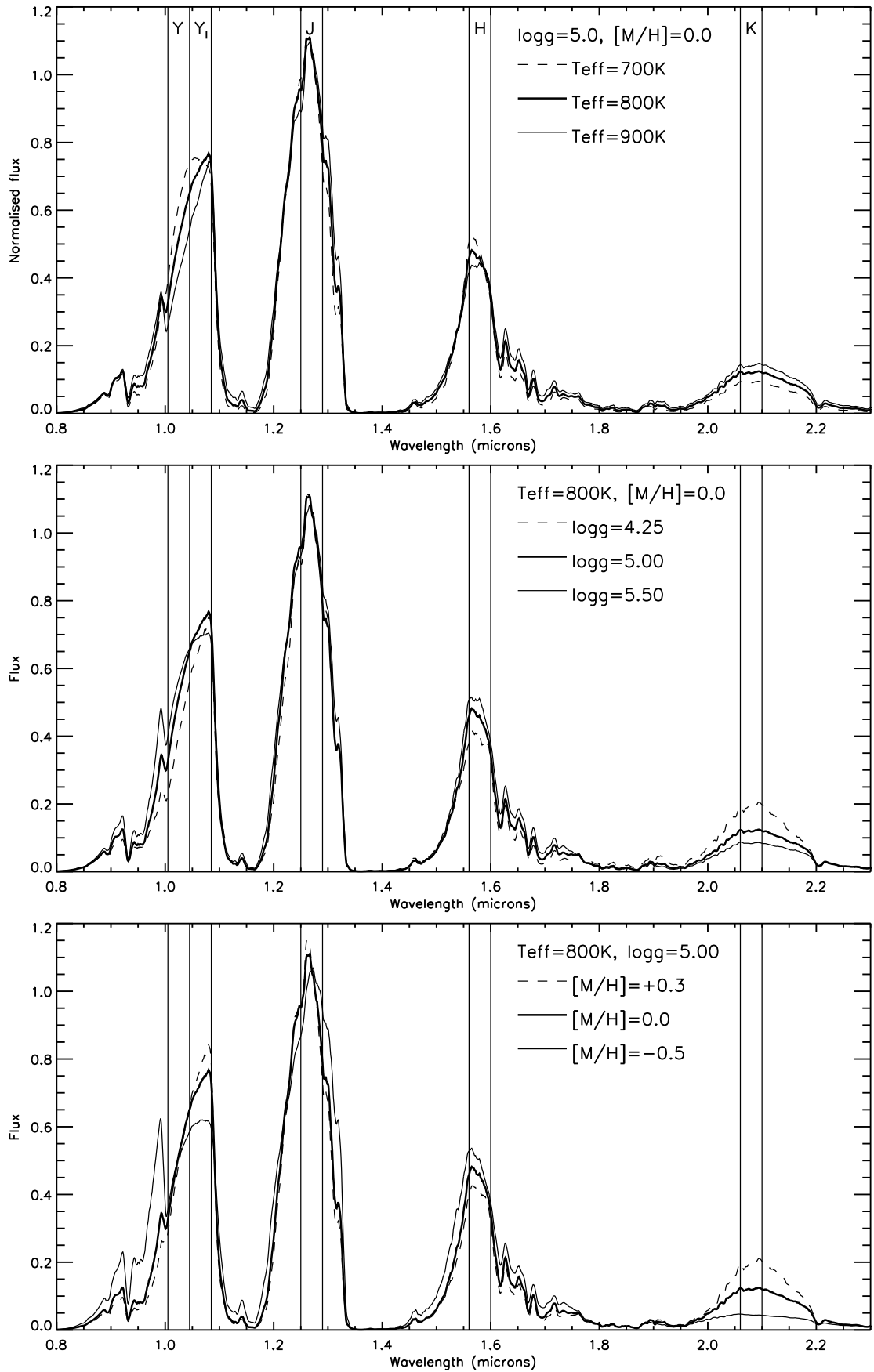}
  \caption{Model BT-Settle T dwarf spectra showing theoretical $T_{\rm eff}$, $\log{g}$ 
and [M/H] variations.}
\end{figure*}

Figure 5 demonstrates how the BT-Settl model spectra change with $T_{\rm eff}$, 
gravity, and metallicity. Our baseline properties in this figure are $T_{\rm eff}$=800K, 
$\log{g}$=5.0, [M/H]=0.0, and spectral variations for $\pm$100K, $^{+0.5}_{-0.75}$ dex, 
and $^{+0.3}_{-0.5}$ dex in $T_{\rm eff}$, $\log{g}$ and [M/H] respectively, are shown. 
The spectra have been normalised to unity in the peak of the $J$-band. As has been noted 
previously for the Tucson and AMES models, $T_{\rm eff}$ variations are significant in 
the wings of the $H$-band peak and in the peak of the $K$-band. The BT-Settl models also 
suggest that the $Y$-band peak has some $T_{\rm eff}$ sensitivity, with increasing and 
decreasing $T_{\rm eff}$ suppressing and enhancing the $Y$-band flux (with respect to 
the $J-$band peak) respectively. By comparison the Tucson models also exhibit some $Y$-band 
variation with $T_{\rm eff}$, but in the opposite sense to that of the BT-Settl models, 
and the AMES models show little variation with $T_{\rm eff}$ (Leggett et al. 2007). 
Gravity has a significant effect on the model $K$-band peak, with higher $\log{g}$ 
causing $K$-band suppression and lower $\log{g}$ enhancing the $K$-band flux. Some 
observational evidence may also support this trend, as for example, S Ori 70 (a possible 
T5.5 member of the $\sim$3Myr $\sigma$Orionis cluster; Zapatero-Osorio et al. 2008), and 
PLZJ93 (a possible T3-5 member of the Pleiades open cluster; Casewell et al. 2007) both 
show a strong $K-$band enhancement, which would presumably result from the low surface 
gravity of members of these solar metallicity clusters.

In addition, the BT-Settl $Y$-band flux is suppressed to some extent at lower gravity. 
However, this effect is not predicted by the Tucson or AMES models (with both models 
showing an increase in flux for lower gravity), and this trend is thus ambiguous when 
taking all models into consideration. Metallicity affects the $K$-band peak in a similar 
way to gravity, where increasing and decreasing metallicity causes $K$-band enhancement 
and suppression respectively. Indeed, $K$-band fluxes may be quite sensitive to even 
small metallicity differences, as suggested by Liu, Leggett \& Chiu (2007) when comparing 
the spectra of HD 3651B (Burgasser 2007) and Gl570D. Also, metallicity affects the $Y$-band 
flux peak significantly. This effect is also seen for the other models, although it is 
not clear how the $Y$-band peak should change. The Tucson and AMES models suggest that 
$Y$-band suppression could occur at higher metallicity, although the BT-Settl models 
suggest that some enhancement may occur (at the expense of shorter wavelength flux).

We do not make direct comparison to the model predictions, but instead identify 
T dwarfs whose relative flux peak ratios and/or colours are unusual in the context 
of T dwarf properties typical of the solar neighborhood. We compared the measurements 
of the T dwarfs with previous samples (from Knapp et al. 2004; Golimowski et al. 2004; 
Chiu et al. 2006), and found that most of the sample (eleven out of fifteen) have ratios 
that appear consistent with the majority of solar neighborhood T dwarfs. However, there 
are some notable exceptions.

\begin{itemize}
  \item{ULAS J0222-0024 and ULAS J0938-0011 appear to show $Y_l$ enhancement. While our 
spectral coverage only allowed us to measure the $Y_l/J$ flux ratio of ULAS J0222-0024 
and ULAS J0938-0011, the strongly enhanced flux in the $Y$-peak is suggestive of unusual 
metallicity. However, there is no evidence for an enhanced $Y$ flux from the $Y-J$ colours. 
It is possible that the $Y_l$-band (but not the $Y-$band) could be enhanced for these 
objects. This would be consistent with the BT-Settl model predictions for a high metallicity 
object (see Figure 5), but the $Y$ spectra of these objects are somewhat noisy and more 
spectral coverage would be needed to properly confirm this.}

  \item{ULAS J0958-0039 may show some $Y$-band and $H$-band enhancement, although the 
somewhat noisy nature of this object's spectrum make this result rather tentative. However, 
it is supported by the relatively blue $Y-J$ colour of this object. Although note that 
the $Y_l$/J spectral ratio of this object is not unusual. Consideration of the models 
suggests that such an effect could arise from lower $T_{\rm eff}$ or unusual [M/H], 
but it is not currently clear if either or both of these effects are responsible.}

  \item{ULAS J1150+0949 shows strong $Y$ and $Y_l$ suppression and also an enhanced $K$-band 
flux. The enhanced $K/H$ ratio and relatively red $H-K$ colour is suggestive of low gravity. 
The suppressed $Y$-band flux for this object also points to low gravity when one considers 
the BT-Settl models. However, note that the Tucson and AMES models (which show no significant 
$Y$-band gravity sensitivity) could invoke an increase in metallicity to explain this $Y$-band 
suppression. While it seems simpler to require only one non-typical property to explain 
the unusual flux peak ratios, it is not possible to make firm conclusions at this 
stage, in light of the model ambiguities. ULAS J0859+1010 also has suppressed $Y_l$ flux 
and unusually red $H-K$ colour, and could have similar physical properties to ULAS J1150+0949.}

  \item{ULAS J1303+0016 shows strong $K$-band suppression and very blue $H-K$ and $J-K$ 
colours, but has a normal $Y$-band flux peak. This is consistent with the predicted 
model trends for a high gravity object.}

  \item{Finally, ULAS J1547+0053 shows typical 
$Y_l/J$ and $H/J$ spectral ratios, but has unusually red $J-K$ and $H-K$ colours. 
Like ULAS J1150+0949 this object may have low gravity, although it does not show 
suppressed $Y_l$ flux.}
\end{itemize}

We summarise the above discussion in the last column of Table 6, where we indicate 
possible non-typical properties that might explain the observed spectral morphology 
and colours.

\begin{figure*}
  \includegraphics[width=15cm]{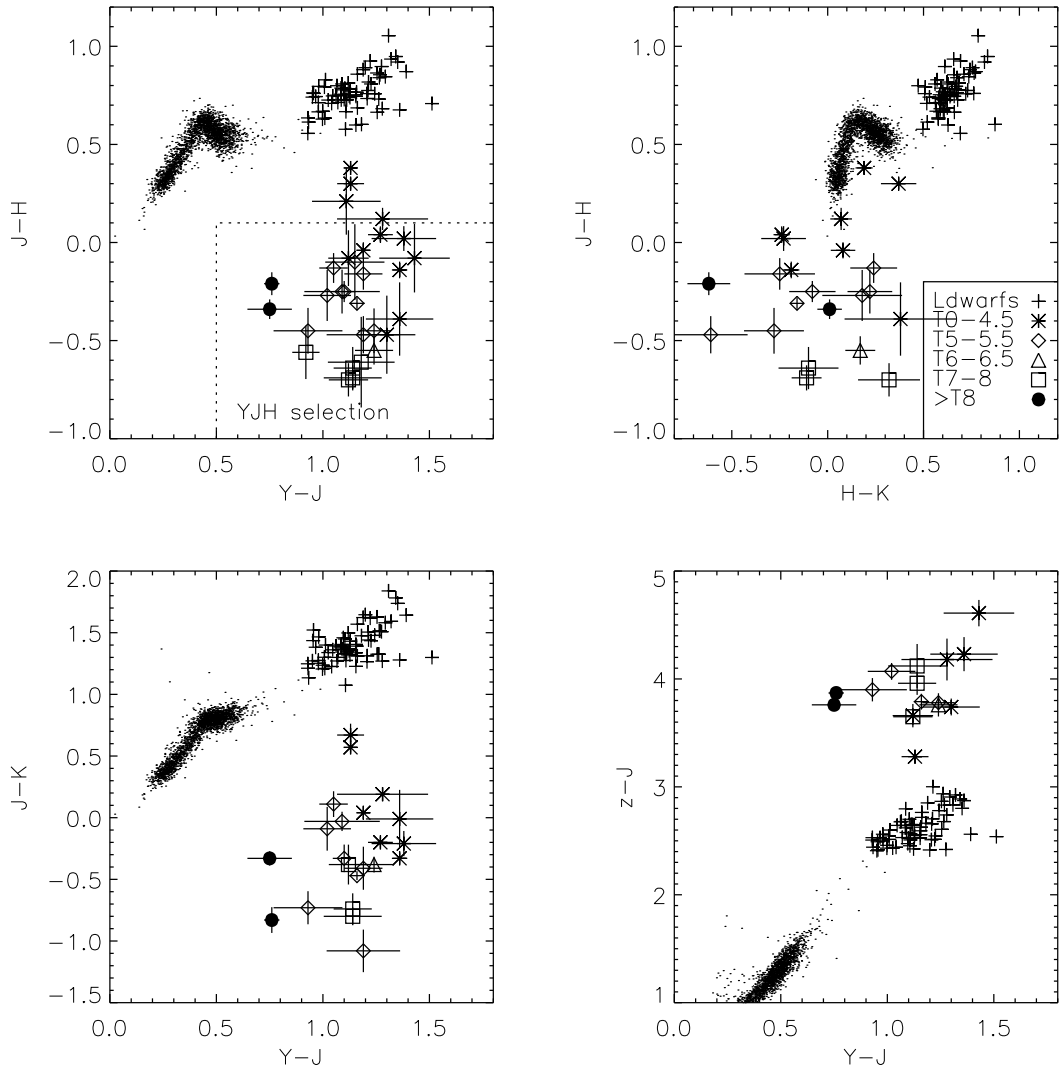}
  \caption{A series of two-colour diagrams showing the known LAS T dwarfs, 
including those first presented by Geballe et al. (2002), Kendall et al. (2007), 
Lodieu et al. (2007), Warren et al. (2007a), Chiu et al. (2007), and this work.}
\end{figure*}

\subsection{Searching for the lowest $T_{\rm eff}$ objects}
Figure 6 shows the available colours of our new T dwarfs as well as other T dwarfs 
that have been discovered or recovered in the LAS (Geballe et al. 2002; Kendall et al. 
2007; Lodieu et al. 2007; Warren et al. 2007a; Chiu et al. 2007), including an 
additional T8+ object CFBDS0059, recently discovered using the Canada-France-Hawaii 
Telescope (Delorme et al. 2008) and independently identified in the LAS. Different 
T dwarf spectral types are indicated with different symbols (see the key in the top 
right plot). Also shown are the sample of DR2 L dwarf candidates (plus signs; see 
Section 2.1), and the sample of typical brighter sources from the LAS (points; see 
section 2.1) for comparison. Our $YJH$ selection box is indicated with a dotted line.

Examination of the plots shows several interesting features. For the blue T dwarfs 
(contained in our selection box) there is some evidence for a possible trend in 
$Y-J$ colour as one moves from the T0-4.5 dwarfs to the T5-8 dwarfs. This trend 
may be continuing into the T8+ objects, as suggested by the rather bluer colour 
of the T8.5 dwarfs ULAS J0034-0052 and CFBDS0059 (filled circles). The $J-H$ colours 
of T dwarfs generally get bluer out to spectral types of T8, although the two T8+ 
dwarfs do not have very blue $J-H$. The $H-K$ and $J-K$ colours show strong variations 
irrespective of spectral sub-class, due presumably to variation in surface gravity 
and metallicity, and such variations appear to continue into the T8+ range. As such, 
these colours may not be strong indicators of T8+ types in themselves, although models 
do indicate that when combined with certain spectroscopic information (e.g. $W_J$ index 
describing the width of the $J$-band peak) such colours (or flux ratios) can provide 
useful $T_{\rm eff}$ constraints (see fig 7 of Warren et al. 2007 and fig 6 of Delorme 
et al. 2008). The $z-J$ colour is clearly rather scattered for T dwarfs, and there is 
certainly no evidence in Figure 6 for T dwarfs becoming redder in $z-J$ at lower 
$T_{\rm eff}$. However, it is interesting to note that the $z-J$ colour of the two 
T8+ dwarfs is perhaps slightly bluer than other mid-late T dwarfs.

Although only two T8+ dwarfs are currently known, their colours suggests that $Y-J$ 
may be a particularly useful colour for identification. The $J-H$ colour shows no 
evidence so far for getting bluer beyond T8, however this is based on only two objects, 
and we cannot rule out the possibility that the small number of T8+ dwarfs discovered 
so far could be unusual. A young (still contracting) brown dwarf for instance, would be 
brighter than an older counterpart of the same $T_{\rm eff}$, and hence detectable 
to greater distance. This could provide an explanation as to why the first discoveries 
in this new $T_{\rm eff}$ range might have non-typical spectral properties. Clearly a 
larger sample of T8+ dwarfs is needed to properly understand their range of spectral 
properties. To this end, it could prove desirable to extend LAS searches for the lowest 
$T_{\rm eff}$ objects to even bluer $Y-J$ colours, and possibly redder $J-H$. Optical 
constraints (e.g. non-detections) would be particularly important as the near infrared 
colours of candidates become more akin to those of hot stars, but the capabilities of 
UKIDSS and SDSS combine to give great potential for such searches.

\section{T dwarf numbers}
In this section we discuss the sample of LAS T dwarfs identified thus far in 
DR1 and DR2, and place a magnitude limited constraint on T dwarf numbers in 
this 280 sq degs of the LAS. We discuss various completeness issues, and make 
approximate corrections to the T dwarf statistics to account for these.

We first consider the completeness with which we have followed up candidates 
from DR1 and DR2. Overall we have followed up 82 per cent of all our $J\le$19.5 
candidates (see Section 3). Nearly all those that still require follow-up are 
fainter than $J$=19. Of the 44 initial candidates with $J\le$19.0, 42 have reliable 
classifications. The follow-up presented here combined with the previous observations 
reported by Kendall et al. (2007), Lodieu et al. (2007), Warren et al (2007a), and 
Chiu et al. (2007), have resulted in the discovery of 23 T dwarfs with $J-H\le$0.1 
in the LAS DR2 sky. Of these, 22 T dwarfs have $J\le$19.0 (10 with $J<$18.5 and 
12 with $J$=18.5-19.0). Since an extra half a magnitude depth should double the 
surveyed volume, it can be seen that our sample shows statistical evidence for 
a good level of completion down to $J$=19.0.

The spread in colour of known T dwarfs means that our colour selection 
($J-H\le$0.1) will thoroughly probe only certain ranges of T spectral class, 
and for some sub-type ranges a fraction of the T dwarfs will be missed. To assess 
the spectral type range that our colour selection probes, we used the population 
of optically selected SDSS T dwarfs\footnote{From the DwarfArchives.org site, 
http://dwarfarchives.org} (to avoid NIR colour selection effects). SDSS T dwarf 
searches reach $z$=20.4, and should thus be complete across the T range 
($z-J\sim$3--4) to $J\sim$16.4. There are currently 19 SDSS T dwarfs with $J<$16.4 
and $J-H_{MKO}>$0.1 which have been found in 6600 sq degs of sky (see Chiu et al. 
2007 and references therein). Amongst these, 14 have spectral type T0--2 and five 
have spectral type T2.5--3.5. None have a spectral type of T4 or greater. Correcting 
these numbers to a LAS $J$=19 depth and sky coverage reveals that although we would 
not expect to have missed a significant number of $\ge$T4 dwarfs, we would expect 
to have missed $\sim$21 T0-2 dwarfs and approximately eight T2.5--3.5 dwarfs. 
Clearly our search should be essentially complete for spectral types $\ge$T4, 
but a correction would be necessary if considering spectral type ranges that 
include earlier T dwarfs. Table 7 shows how this correction factor changes for 
different ranges of T spectral class, and how the over-all expected numbers 
of T dwarfs would be affected.

To take advantage of our good level of completeness and facilitate simple comparison 
with the theoretical predictions using the methods of Deacon \& Hambly (2006), we 
specifically consider the spectral type range T4 and later. Allowing for the two 
outstanding candidates we place a constraint of 19--21 T4-T8.5 dwarfs with measured 
$J-H\le$0.1 and $J\le$19 in the LAS Dr2 sky, and allow for a additional 10\% 
Poisson uncertainty on the upper limit to account for the limited statistical 
accuracy of our SDSS sample analysis.

To account for measurement uncertainties, we have estimated the number of $\ge$T4 
dwarfs with intrinsic $J-H\le$0.1 whose colour may have been statistically scattered 
out of the selection region. To do this we estimated (for each $\ge$T4 dwarf discovered 
via $YJH$ detection) the probability that $J-H$ colours could have been scattered 
sufficiently red-ward from the LAS measured value, taking into account the 1-$\sigma$ 
uncertainties on their measured colour. Probabilities were then summed, and our 
result suggests that we would only expect $\sim$1.5 $\ge$T4 dwarfs to be missed 
in this way.

In addition, we account for some level of unresolved binarity amongst the LAS 
T dwarfs. Unresolved binaries will be brighter than a single T dwarf population, 
and could thus be found in a larger volume of sky. A correction is necessary if one 
wishes to compare T dwarf numbers with theoretical predictions for the total number 
of individual T dwarfs. Brown dwarf binaries are generally tight systems (e.g. $<$15AU; 
Reid et al. 2006), of approximately near equal mass components. A volume limited 
sub-stellar binary fraction (BF) has been reported with values ranging from 10--50 
per cent via open cluster photometry and high resolution imaging (e.g. Burgasser 
et al. 2003; Pinfield et al. 2003; Lodieu et al. 2007), where 

\begin{equation}
BF = \frac{N_b}{N_b + N_s}, 
\end{equation}

$N_b$ and $N_s$ being the number of binary and single systems respectively. If 
one makes the simplifying assumption of binary components of equal brightness 
(see Burgasser et al. 2003 for a more generalised conversion for a range of 
mass-ratio distributions; e.g. Allen 2007), then the relative number of binary 
and single systems in a magnitude limited sample will be 

\begin{equation}
\frac{N_b}{N_s} = \frac{2\sqrt{2} BF}{1-BF}, 
\end{equation}

and the number of binary systems (compared to the total magnitude limited number 
of systems $N_m$) will be 

\begin{equation}
N_b = \left( \frac{2\sqrt{2}}{2\sqrt{2}+\frac{1}{BF}-1} \right) N_m.
\end{equation}

For BF=10--50\%, we thus estimate that 24--74 per cent of the sources in our magnitude 
limited sample could be unresolved binaries. Of these unresolved binaries systems, only 
$\sim$35 per cent ($1/2\sqrt{2}\times$100\%) would be included in our $J\le$19 sample 
if the binary components were resolvable (i.e. if the combined binary $J$-band brightness 
is $>$18.25 then the individual T dwarf components would have $J>$19 and would be excluded 
from our sample). However, for the $J\le$18.25 unresolved binary systems, we must count 
both of the T dwarf components. We thus derive a binarity correction factor of 0.76--0.93 
(for BF=10--50 per cent respectively) to correct our numbers to represent a magnitude 
limited sample of individual T dwarfs. Note that despite a wide range in BF, this binary 
correction factor has a relatively small range.

We also account for spatial incompleteness which can result from the way in which 
optical catalogues are used to inform candidate selection. In our initial selection, 
a nearby optical source might result in a candidate being rejected if it was close 
enough to mascarade as an optical counterpart, making the candidate appear too bluer 
in its optical-infrared colour. This effectively decreases the search area by some 
fraction that is dependent on the source density and the search criteria used. We 
account for this with a correction factor ($\epsilon$) that is the ratio of clear 
sky (unaffected by optical mis-matches) to total sky considered ($A_{tot}$), and 
can be derived from probability analysis as, 

\begin{equation}
\epsilon = \left(1-\frac{\sigma}{A_{tot}}\right)^N, 
\end{equation}

where $\sigma$ is the area excluded by a single optical source and $N$ is the number 
of optical sources in the area considered. For 1 deg$^2$ this reduces to 
$\epsilon = (1-\sigma)^n$ where $n$ is the number density of optical sources. 
Figure 7 shows $\epsilon$ plotted against source density for our search of the 
LAS (solid line) as well as for previous searches of 2MASS (dashed line, e.g. 
Burgasser et al. 2004). The differences result from the different exclusion regions 
used, which extended out to 10 arcsecs from any USNO-A2.0 optical source for 2MASS 
searches, but only extend out to 2 arcsecs from any SDSS source for our search of the 
LAS. The full effect of this incompleteness depends on the number density of optical 
sources in the regions of sky being considered, and this in turn is dependent on 
the galactic latitude. LAS DR2 extends across the $b=$20-80 deg range, although is 
mostly concentrated in $b=$40-75 deg. The SDSS source number density ranges from 
$\sim$20,000-40,000 sources per square degree over the LAS DR2 sky (indicated by open 
diamonds in Figure 7), and as can be seen, this results in a range of $\epsilon$=0.96-0.98. 

\begin{figure}
  \includegraphics[width=8cm]{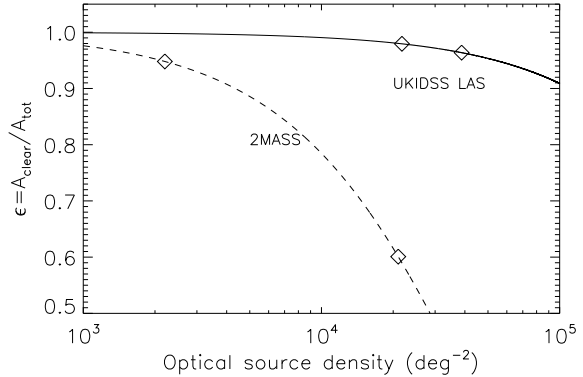}
  \caption{Predicted area corrections, $\epsilon = \frac{A_{clear}}{A_{tot}}$ for 
excluding regions around optical sources as a function of source density. The UKIDSS 
LAS exclusion regions comprise 2 arcsec radii circles centred on SDSS sources, and 
for comparison the 2MASS exclusion regions comprise 10 arcsec radii circles centred 
on USNO sources. The source densities inherent in our search and in previous 2MASS 
searches out of the plane are enclosed by open diamonds.}
\end{figure}

For the 2MASS comparison, despite the optical surveys being shallower than SDSS, 
the larger exclusion regions (with 10 arcsec radii) combined with the greater range 
of galactic latitude searched (reaching $|b|=15$) result in the $\epsilon$ correction 
having a potentially much greater impact than it does for the LAS DR2 T dwarf sample. 
We chose to factor in a 3\% increase to our T dwarf numbers to account for this 
spatial incompleteness effect, but note that this small correction has no significant 
effect on our results.

Finally, we account for Malmquist bias. We first consider the effect of an intrinsic 
scatter in the M$_J$ of T dwarfs (classical Malmquist bias). This scatter will cause 
the $J$-band magnitude limited sample to probe different volumes for any fixed spectral 
type. Those T dwarfs that are intrinsically brighter than average (for their type) will 
be sampled from a larger volume than those that are fainter. This will bias the mean 
M$_J$ to brighter values, and effectively result in the overall sample coming from a 
larger volume than if the T dwarfs had a non scattered M$_J$-spectral type relation. 
Assuming a uniform space density and ignoring any effects of slope in the substellar 
luminosity function (i.e. $d\phi/d{\rm M}=0$), the fractional increase in the measured 
number of T dwarfs ($\phi$) is given by 

\begin{equation}
\frac{\Delta\phi}{\phi} = \left(\frac{0.6\sigma}{\log{e}}\right)^2
\end{equation}

where $\sigma$ is the rms dispersion (assuming a Gaussian distribution) of T dwarf 
absolute magnitudes. We estimated this scatter using the sample of 
T dwarfs with parallax measurements presented by Liu et al. (2006) combined with 
their polynomial fit (M$_J$-spectral type) to the single T dwarfs (ignoring known 
and possible unresolved binaries). When one considers only the subset of T dwarfs 
with spectral type $\ge$T4 (i.e. matched to the DR2 sample), the rms scatter from 
the best-fit polynomial is $\sigma$=0.29, resulting in an increased volume of 
$\Delta\phi/\phi$=16\%. We thus decreased our T dwarf numbers to account for 
this bias.

An additional Malmquist effect results from photometric uncertainty near the 
sample limit, since more T dwarfs scatter into the sample from fainter magnitude 
(i.e. at a greater distance, in a larger volume) than scatter out of it. However, 
since we measured accurate follow-up photometry for the majority of 
our initial candidates (down to $J$=19.5) before we imposed the $J$=19.0 cut 
considered here, the effective scatter at the $J$=19 limit will be small. We 
conservatively estimate this scatter to be $\sigma$=0.05, and expect a resulting 
Malmquist correction of no more than 0.5\% which is insignificant.

Our final magnitude limited T dwarf number constraints take account of the above 
corrections, and assumes Poisson uncertainties associated with sample sizes. We thus 
estimate that there are 17$\pm$4 single $\ge$T4 dwarfs with $J<$19 in the 280 sq degs 
of Dr2 sky. Table 7 also presents the number constraints derived using different lower 
limits for the spectral type range.

\begin{table*}
\centering
\caption{T dwarf sample statistics.}
\begin{tabular}{lccc}
\hline
Spectral   & No. in our  & Correction to    & Final DR2 \\
type range & $J<19$ DR2  & include T dwarfs & corrected \\
           & sky (280 sq & with $J-H>$0.1   & number to \\
           & degs)       &                  & J=19 $^a$ \\
\hline
  T0--T8.5 & 22--24     &  +95$\pm$27\%      & 35$\pm$9 \\
T2.5--T8.5 & 22--24     &  +40$\pm$15\%      & 24$\pm$6 \\
  T4--T8.5 & 19--21     &       $<$10\%      & 17$\pm$4 \\
\hline
\multicolumn{4}{l}{$^a$Numbers corrected to account for redder ($J-H>$0.1) T dwarfs}\\
\multicolumn{4}{l}{outside our selection, LAS photometric uncertainties, unresolved}\\
\multicolumn{4}{l}{binarity for a binary fraction in the 10--50\% range, spatial}\\
\multicolumn{4}{l}{incompleteness and Malmquist bias (see text).}\\
\hline
\end{tabular}
\end{table*}

For comparison, we undertook a series of simulations based on those presented in 
Deacon \& Hambly (2006). These were scaled to take into account the difference in 
size between the full LAS and DR2. The $J$-band depth limit was set to 19, and 
the $Y$-band depth limit to 20. This $Y$-band depth limit is actually 0.2 magnitudes 
brighter than the 5$\sigma$ detection limit, but this brighter limit effectively 
takes into account the known drop in detection completeness for J=19.7--20.2 (Irwin 
priv. comm.). Unlike the Deacon \& Hambly (2006) simulations we set no limit on the 
$H$-band detection, since objects not detected in the $H$-band will still be identified 
as part of our $YJ$ selection. Also, note that the Deacon \& Hambly (2006) simulations 
used a normalisation value of 0.0055 objects per cubic parsec in the mass range 
0.1-0.09$M_\odot$ taken from Burgasser (2004). Here we use a normalisation value of 
0.0038$\pm$0.0013 per cubic parsec calculated by Deacon, Nelemans and Hambly (2008). 
The results of these simulations are shown in Table 8, where the simulated late 
T dwarfs were defined as having $T_{\rm eff}<1300K$. Mass function power-law 
indices $\alpha$ and $x$ are also indicated, where the simulations assume functions 
of the form $dn/dm\propto$m$^{-\alpha}$ or $dn/d(\log{m})\propto$m$^{-x}$ (following 
Chabrier 2005).

\begin{table}
\centering
\caption{Simulated late T dwarf (defined as those with $T_{\rm eff}<$1300K) numbers 
for LAS DR2 (see text), for which we assumed input forms for the initial mass function 
of $dn/dm\propto m^{-\alpha}$ or $dn/d(\log{m})\propto m^{-x}$, and for the birth-rate 
of $b(t)\propto e^{-\beta t}$.}
\begin{tabular}{llc}
\hline
\multicolumn{3}{l}{Constant birthrate}\\
\hline
$\alpha=-1.0$ & $x=-2.0$ &  15$\pm$5  \\
$\alpha=-0.5$ & $x=-1.5$ &  25$\pm$9  \\
$\alpha=0.0$  & $x=-1.0$ &  44$\pm$15 \\
$\alpha=0.5$  & $x=-0.5$ &  57$\pm$19 \\
$\alpha=1.0$  & $x=0.0$  & 119$\pm$40 \\
\hline
\multicolumn{3}{l}{IMF with $\alpha=0.0$ or $x=-1.0$}\\
\hline
$\beta=-0.1$ & &   54$\pm$18 \\
$\beta=0.0$  & &   44$\pm$15 \\
$\beta=0.1$  & &   33$\pm$11 \\
$\beta=0.2$  & &   35$\pm$12 \\
$\beta=1.0$  & &   31$\pm$11 \\
\hline
\end{tabular}
\end{table}

When comparing the observed number of T4-T8.5 dwarfs (from Table 7) to the 
predicted numbers in Tables 8, and allowing for the associated uncertainties, 
it is clear that it is not possible to place a significant constraint on the birthrate. 
However, this comparison favours a range of $\alpha$ between -0.5 and -1.0. 
Statistically our T dwarf numbers are also reasonably consistent (to within 1.5$\sigma$) 
with $\alpha=0.0$. However, they are inconsistent with $\alpha$ of 0.5 and 1.0 at the 
2.0 and 2.5$\sigma$ level respectively. This result is reasonably consistent with the 
finding of Metchev et al. (2008), who performed a combined 2MASS and SDSS T dwarf 
search and derived a T dwarf space density that was most consistent with $\alpha=0.0$ 
(although based on lower number statistics than here). However, L dwarf mass function 
constraints suggest higher values of $\alpha$ in the 1-1.5 range (Reid et al. 1999; 
Cruz et al. 2007). The sub-stellar mass function of young open clusters are generally 
consistent with $\alpha\simeq0.5$. However, a lognormal function also offers a reasonable 
fit to observation (see the Pleiades mass function in figure 4 of Chabrier 2005). The 
slope of the lognormal mass function gets steeper as one decreases brown dwarf mass, 
and for $\sim$0.04$M_{\odot}$ the lognormal mass function slope is consistent with an 
$\alpha=0.0$ power law. A detailed comparison between mass function constraints from 
L and T dwarfs is beyond the scope of this paper, however, it may be that differences 
occur due to the T dwarfs probing a somewhat lower mass range than the L dwarfs. This 
would suggest that a single power-law exponent is not optimal when describing the 
substellar mass function in the field. Clearly a greatly improved picture of the 
mass function will emerge from the LAS as the survey area grows.

\section{Conclusions and future work}
We have discussed our database selection methods and the photometric follow-up 
that we performed to search for late T dwarfs and even cooler $T_{\rm eff}$=400--700K 
objects in the UKIDSS LAS. These techniques have allowed us to make essentially 
complete follow-up of our candidates from DR1 and DR2 down to $J$=19. We have also 
followed up some candidates from DR3 and a fraction of our fainter candidates from 
DR1 and DR2 (to J=19.5). Using a variety of spectroscopic facilities we have measured 
the spectra of our best candidates, and have spectroscopically confirmed 15 new T dwarfs, 
bringing the total of number of confirmed LAS T dwarf discoveries to 28.

Compared to typical T dwarf properties, one of our new T dwarfs may be metal-poor, 
two may have relatively low surface gravity, one may have higher than normal 
surface gravity, and one may be metal rich or have low surface gravity. These assessments 
are based on comparisons between the spectral morphology and colours of the T dwarfs 
with theoretical models trends, and are thus somewhat speculative. However, all indications 
are that T dwarf spectra are quite sensitive to their physical properties, and it seems 
clear that a better understanding of these variations would have important implications 
for our ability to model T dwarf atmospheres, and potentially constrain (via spectroscopy) 
the mass, age and composition of T dwarf populations in the field (e.g. Pinfield et al. 2006). 
To this end, it would be very beneficial to identify T dwarfs whose properties could be 
constrained without the need to study their spectra. Both T dwarfs and L dwarfs of this 
type may be found in widely separated binary systems (e.g. Gizis et al. 2001; Wilson et al. 
2001; Pinfield et al. 2006; Scholz et al. 2003; Burgasser et al. 2000; Luhman et al. 2007) 
for which the primary object can be used to constrain system age and composition. Gl570D 
and HD3651B (Burgasser 2000 and Luhman et al. 2007) are good examples, being analysed for 
age and abundance by Geballe et al. (2001), Burgasser (2007) and Liu Leggett \& Chiu (2007).

The large volume (and large number of T dwarfs) probed by the LAS should yield 
many T dwarfs in such binary systems, and in this context we made a basic search for 
possible companions to all confirmed LAS T dwarfs out to a separation of $\sim$10,000AU 
(at the estimated distances in Table 5), by querying the Simbad astronomical database 
(operated at CDS). We searched for neighboring objects with either a measured spectral 
type, a parallax, or a high proper motion, for which we could obtain distance constraints 
from either parallax or from spectral type and photometry. These distance constraints 
allowed us to rule out (as possible companions) all the neighboring objects, by 
comparison with the distance constraints of the T dwarfs. However, proper motion 
measurements for all LAS T dwarfs would facilitate a more general search for common 
proper motion companions as the number of LAS T dwarfs grows, which could include 
white dwarf companions that can yield useful age constraints (e.g. Day-Jones et al. 
2008) and are readily identified in SDSS (e.g. Kleinman et al. 2004; Gates et al. 2004; 
Eisenstein et al. 2006).

The number of blue ($J-H<$0.1) LAS T dwarfs discovered with our methods shows statistical 
evidence for a good level of completeness down to $J$=19. This builds on our previous work 
(e.g. Lodieu et al. 2007) for which a good level of completion was only achieved to $J\sim$18.5. 
The increased size of the LAS T dwarf sample allowed us to place some statistical constraints 
on the sub-stellar mass function. Indeed, there may be mounting evidence for a steepening 
(decreasing more rapidly with decreasing mass) sub-stellar mass function in the field, with 
best-fit $\alpha$ values of $\sim 1$ and between -1.0 and 0.0 for L and T dwarf populations 
respectively. In any event, mass function constraints should be greatly improved by larger 
samples of LAS T dwarfs as the survey continues to grow. In addition, constraints on the brown 
dwarf birth-rate should be attainable by building on the LAS L and early T dwarf searches mentioned 
briefly in Section 2.1 (e.g. fig 3 of Deacon \& Hambly; fig 10 of Burgasser 2004). For instance, 
a sample of $\sim$100 L/T dwarfs per $\Delta T_{\rm eff}$=100K range from 1100-1500K (assuming 
a flat birth-rate) would be capable of ruling out an exponential (e.g. $\tau_g\sim$5Gyr) 
birth-rate at a $\sim$5$\sigma$ level of significance. An appropriate sample of $\sim$400 
late L to mid T dwarfs could be identified in the full LAS to $J$=18.5 (for a flat 
birth-rate).

The discovery of the T8.5 dwarf ULAS J0034-0052 (Warren et al. 2007) in the LAS absolutely 
demonstrates that UKIDSS is probing unexplored $T_{\rm eff}$ regimes beyond the previously 
known late T dwarfs. Having said this, parallax and adaptive optics measurements for this 
and other very late T dwarfs are important if we are to test for unresolved binarity and 
better constrain luminosity and $T_{\rm eff}$. When comparing the colours of ULAS J0034-0052 
with all available LAS T dwarfs, it is possible that a blue trend in $Y-J$ with 
decreasing T$_{\rm eff}$ may be present. The $J-H$ and $z-J$ colours might also change 
beyond T8, becoming slightly redder and bluer respectively, although this needs further 
investigation. These spectral changes would be of great importance not only to our 
understanding of cool dwarf atmospheres, but also to the potential for finding such 
objects in the LAS. If it is the case that $T_{\rm eff}$=400-700K objects become 
bluer in $Y-J$ and redder in $J-H$ than T8 dwarfs, it could be less problematic to 
identify such candidates to greater $J-$band depth in multi-band searches of the 
combined LAS and SDSS databases. Indeed, an evolving set of search criteria takes 
full advantage of the capabilities offered by large general surveys like UKIDSS and 
SDSS.

\section*{Acknowledgments}

This work is based in part on data obtained as part of the UKIRT Infrared Deep 
Sky Survey. The United Kingdom Infrared Telescope is operated by the Joint 
Astronomy Centre on behalf of the Science and Technology Facilities Council 
of the U.K. Some of the data reported here were obtained as part of the UKIRT 
Service Programme. Based on observations obtained at the Gemini Observatory, 
which is operated by the Association of Universities for Research in Astronomy, 
Inc., under a cooperative agreement with the NSF on behalf of the Gemini partnership: 
the National Science Foundation (United States), the Science and Technology 
Facilities Council (United Kingdom), the National Research Council (Canada), 
CONICYT (Chile), the Australian Research Council (Australia), CNPq (Brazil) 
and SECYT (Argentina). Based in part on data collected at Subaru Telescope, 
which is operated by the National Astronomical Observatory of Japan. SKL's research 
is supported by the Gemini Observatory. Based on 
observations made with the Italian Telescopio Nazionale Galileo (TNG) operated 
on the island of La Palma by the Fundacin Galileo Galilei of the INAF (Istituto 
Nazionale di Astrofisica) at the Spanish Observatorio del Roque de los Muchachos 
of the Instituto de Astrofisica de Canarias. This publication makes use of data 
products from the Two Micron All Sky Survey, which is a joint project of the 
University of Massachusetts and the Infrared Processing and Analysis Center/California 
Institute of Technology, funded by the National Aeronautics and Space Administration 
and the National Science Foundation. Funding for the SDSS and SDSS-II has been 
provided by the Alfred P. Sloan Foundation, the Participating Institutions, the 
National Science Foundation, the U.S. Department of Energy, the National Aeronautics 
and Space Administration, the Japanese Monbukagakusho, the Max Planck Society, and 
the Higher Education Funding Council for England. The SDSS Web Site is http://www.sdss.org/. 
The SDSS is managed by the Astrophysical Research Consortium for the Participating 
Institutions. The Participating Institutions are the American Museum of Natural History, 
Astrophysical Institute Potsdam, University of Basel, University of Cambridge, Case Western 
Reserve University, University of Chicago, Drexel University, Fermilab, the Institute for 
Advanced Study, the Japan Participation Group, Johns Hopkins University, the Joint Institute 
for Nuclear Astrophysics, the Kavli Institute for Particle Astrophysics and Cosmology, the 
Korean Scientist Group, the Chinese Academy of Sciences (LAMOST), Los Alamos National 
Laboratory, the Max-Planck-Institute for Astronomy (MPIA), the Max-Planck-Institute for 
Astrophysics (MPA), New Mexico State University, Ohio State University, University of 
Pittsburgh, University of Portsmouth, Princeton University, the United States Naval 
Observatory, and the University of Washington. The William Herschel Telescope is 
operated on the island of La Palma by the Isaac Newton Group in the Spanish 
Observatorio del Roque de los Muchachos of the Instituto de Astrofisica de 
Canarias. This research has benefited from the M, L, and T dwarf compendium 
housed at DwarfArchives.org and maintained by Chris Gelino, Davy Kirkpatrick, 
and Adam Burgasser. This research has made use of the SIMBAD database, operated 
at CDS, Strasbourg, France.

\bsp

\label{lastpage}

\end{document}